%% file: Manuscript.tex
\documentclass[oneside,reqno,11pt,a4paper]{amsart}
\usepackage[T1]{fontenc}
\usepackage[margin=2.5cm]{geometry}
\usepackage{bm}
\usepackage{amsmath,amsfonts,amssymb,amscd,amsthm}
\usepackage[usenames,dvipsnames,svgnames]{xcolor}
\usepackage{graphicx}
\graphicspath{}
\usepackage{caption}
\usepackage{subcaption}
\usepackage{doi}
\usepackage{makecell}
\usepackage{hyperref}
\usepackage{acronym}
\usepackage{listings}
\usepackage{textgreek}
\usepackage{url}
\usepackage{color}
\usepackage{framed}
\usepackage{verbatim}
\usepackage{fancyvrb}
\usepackage{microtype} 
\usepackage[final]{pdfpages}
\usepackage{tikz}
\usepackage{booktabs}
\usepackage{multirow}
\usepackage{algorithm}
\usepackage{algpseudocode}

\usepackage[backend=biber,
defernumbers=true,
maxbibnames=5,
style=numeric-comp,
isbn=false,
bibencoding=utf8,
safeinputenc, % https://github.com/plk/biblatex/issues/819
url=false,
doi=true,
giveninits=true,
sorting=none,
]{biblatex}

\usepackage{textcomp}
\usepackage{cancel}
\usepackage{mathrsfs}
\usepackage{stmaryrd}
\usepackage{parskip} %stopping indentation
\usepackage{import}
\usepackage{xifthen}
\usepackage{pdfpages}
\usepackage{hyperref}
\usepackage{float}
\usepackage{graphicx}
\usepackage{siunitx}
\usepackage{placeins}
\usepackage[labelfont=bf]{caption}
\usepackage{enumitem}
\usepackage{cleveref}
\usepackage{import}
\usepackage{transparent}
\usepackage{xcolor}
\usepackage{longtable}
\catcode`\|=12\relax
\usepackage{pgfplots}
\pgfplotsset{compat=newest}
\usepackage{program}
\usepackage{tikzscale}
\usepackage{soul}
\usepackage{pmboxdraw}

\tikzset{
every picture/.style={
line width = 0.3mm} %or use: "`line width=1 pt,"<-- note:if you write line width, you must use a value with unit
}
\pgfplotsset{
every axis/.append style={
line width = 0.3mm,
grid style={
    line width = 0.3mm,
},
tick style={
    line width = 0.3mm,
},
},
}

\makeatletter
\let\blx@rerun@biber\relax
\makeatother

\hypersetup{
breaklinks=true,
bookmarksopen=true,
pdftitle={Template article},    % insert title
pdfauthor={BadiaLab},     % insert author
colorlinks=true,       % false: boxed links; true: colored links
linkcolor=black,          % color of internal links (change box color with linkbordercolor)
citecolor=blue,        % color of links to bibliography
filecolor=black,      % color of file links
urlcolor=blue           % color of external links
}
\definecolor{bg}{rgb}{0.93,0.93,0.93}

\acrodef{pde}[PDE]{partial differential equation}
\acrodef{fe}[FE]{finite element}
\acrodef{fem}[FEM]{finite element method}
\acrodef{fcm}[FCM]{finite cell method}
\acrodef{DOF}[DOF]{degrees of freedom}
\acrodef{agfem}[AgFEM]{aggregated finite element method}
\acrodef{cutfem}[CutFEM]{cut finite element method}
\acrodef{ls}[LS]{level set}
\acrodef{to}[TO]{topology optimization}
\acrodef{nn}[NN]{Neural network}

\newcommand{\tnor}[1]{{\left\vert\kern-0.25ex\left\vert\kern-0.25ex\left\vert #1 
\right\vert\kern-0.25ex\right\vert\kern-0.25ex\right\vert}}

\newcommand{\sbcom}[1]{{{{#1}}}}

\begin{document}

\newcommand{\incfig}[2][1]{%
    \def\svgwidth{#1\linewidth}
    \import{./figures/}{#2.pdf_tex}
}

\newcommand{\mycomment}[1]{}

\renewcommand{\multicitedelim}{\addcomma}

\title[Neural Level Set Topology Optimization Using Unfitted Finite Elements]{Neural Level Set Topology Optimization Using unfitted Finite Elements}


\author[]{Connor N. Mallon$^{1*}$}
\author[]{Aaron W. Thornton$^{2}$}
\author[]{Matthew R. Hill$^{1,2}$}
\author[]{Santiago  Badia$^{3*}$}
\thanks{\null\ 
$^{1}$ Department of Chemical and Biological Engineering, Monash University, Wellington Rd Clayton, 3800, Victoria, Australia.\ 
$^{2}$ CSIRO, Research Way Clayton, 3168, Victoria, Australia,\ 
$^{3}$ School of Mathematics, Monash University, Wellington Rd Clayton, 3800, Victoria, Australia.\ 
$^*$ Corresponding authors\ 
E-mails: {\tt connor.mallon@monash.edu} (Connor Mallon, Department of Chemical and Biological Engineering, Monash University, Wellington Rd Clayton, 3800, Victoria, Australia), {\tt santiago.badia@monash.edu} (Santiago Badia, School of Mathematics, Monash University, Wellington Rd Clayton, 3800, Victoria, Australia)}


\begin{abstract}
To facilitate the widespread adoption of automated engineering design techniques, existing methods must become more efficient and generalizable. In the field of topology optimization, this requires the coupling of modern optimization methods with solvers capable of handling arbitrary problems. In this work, a topology optimization method for general multiphysics problems is presented. We leverage a convolutional neural parameterization of a level set for a description of the geometry and use this in an unfitted finite element method that is differentiable with respect to the level set everywhere in the domain. We construct the parameter to objective map in such a way that the gradient can be computed entirely by automatic differentiation at roughly the cost of an objective function evaluation. Without handcrafted initializations, the method produces regular topologies close to the optimal solution for standard benchmark problems whilst maintaining the ability to solve a more general class of problems than standard methods, e.g., interface-coupled multiphysics.
\end{abstract}

\maketitle

\input{litreview.tex}

\input{inverseproblem.tex}

\input{network.tex}

\input{lsprocessing.tex}

\input{FCM.tex}

\input{derivative.tex}

\input{results.tex}

\input{conclusion.tex}

\printbibliography  

\end{document}

%% file: litreview.tex
\section{Introduction}\label{introduction}
After the birth of \acp{to} in the field of structural design \cite{Bendse1989}, efforts have been made to increase the effectiveness of such automated design approaches and allow for their deployment on a more general class of problems \cite{Sigmund2013,Guo2010}.   

A plethora of \ac{to} strategies exist through the literature, the most common of which being density-based methods using the so-called SIMP (Solid isotropic microstructure with penalization for intermediate densities) method \cite{Sigmund2001}. These involve varying a material distribution continuously 
between 0 and 1 to introduce an artificial representation of the
boundary. Although simple for basic structural problems, a way to represent intermediate design variables arising at
the boundary must be included, which becomes increasingly
complex in multiphysics applications and makes imposing
arbitrary boundary conditions non-trivial \cite{Yoon2014}. 

%para 4 : \ac{ls} topopt
An alternative technique that can overcome some of the problems presented by density methods and tackle a more general class of problems (e.g., interface-coupling multiphysics and problems that involve surface PDEs on boundaries) is the \ac{ls} \ac{to} method \cite{Osher1988,SethianJamesAlbert1999Lsma}. Using this approach, the boundary is
described by the zero iso-surface of an \ac{ls} function. It is instead this \ac{ls} function that is
varied to obtain optimized designs. A precise location of the boundary
is then available. 

A variety of alternative implementations of the \ac{ls} \ac{to} method
have been made \cite{vanDijk2013}. They can be distinguished, among other things, by how they update the topology at each iteration and their means of geometry mapping.
The methods to update geometries involve either updating the solution of Hamilton-Jacobi equations by a velocity field based on sensitivity information \cite{Osher2001, Burman2018} or using a parameterization of the topology that is an explicit function of the design variables of a steepest descent optimization scheme. The latter approach allows one to leverage well-established nonlinear programming techniques and is the method selected for this work.

Types of geometry mappings include using the \ac{ls} function to define a conformal mesh to the boundary (see e.g. \cite{Ha2008,Yamasaki2011}) which requires re-meshing at each iteration, 
density-based mappings (see e.g. \cite{Allaire2004,Wang2003,Dugast2020}), which recover some of the issues related to density methods, 
or unfitted/immersed boundary techniques (see, e.g., \cite{Parvizian2011,Burman2015,badia_stokes_2018}). Unfitted methods rely on a fixed background mesh and capture the precise location of the boundary using triangulations of the cells cut by the \ac{ls} function. By doing so, re-meshing is avoided yet an accurate description of the interface is maintained. \sbcom{In contrast to density-based mappings, or ersatz material approaches, boundary conditions other than Neumann conditions can easily be imposed on the precise location of the interface. This is critical for the handling of multiphysics problems where one must impose transmission conditions. Furthermore, the sharp treatment of the boundary in the method also means that there is no integration error on any interface or boundary. }

A known issue with unfitted techniques is the ill-conditioning problem associated with small cut elements. The common XFEM \cite{Kreissl2012,Villanueva2017} approach uses a \ac{fe} space restricted to the interior domain and cut cells for the solution and requires stabilization in the vicinity of the boundary by, for example, ghost penalty terms \cite{Burman2010} or cell aggregation \cite{badia_aggregated_2017,Badia2022-linking}. These methods are consistent and can provide high-order approximation \cite{Badia2022-high} however the support of the stabilization terms changes depending on the location of the cut cells, leading to potential non-differentiability in the optimization problem which can harm the convergence of gradient-based optimization algorithms. The specific unfitted \ac{to} technique used in this work is instead a version of the \ac{fcm} \cite{Parvizian2011}, in which a non-consistent penalty term is added everywhere in the fictitious domain (outside the physical domain) to provide robustness. This stabilization is suitable for \ac{to} because it is differentiable with respect to the level set parameterization (see Section \ref{simulation}). An implementation of the \ac{fcm} for \ac{to} is made in \cite{Parvizian2011}, which uses a refined grid for the material boundary compared to the solution to capture fine-scale geometry. We instead use subgrid triangulations using the \ac{ls} function as in \cite{Kreissl2012} to capture fine-scale structure in the integration and thus avoid the need to increase the number of design variables parameterizing the geometry. The loss of consistency of the \ac{fcm} is not an issue in TO, where high-order approximations are not very relevant. 

\sbcom{When utilizing nonlinear programming techniques, the user is free to select a particular parameterization of the geometry. With a mesh already defined for the FE problem,
it is natural to also use a \ac{fe} function for a discrete representation of the \ac{ls}. Doing so, a parameterization is obtained with a user-controlled resolution. A common approach to the optimization problem is then to take the \ac{DOF} values of this \ac{fe} function as the design variables \cite{Kreissl2011,Kreissl2012,Dijk2012}. This choice, however, means that each parameter is only capable of a local influence on the geometry. This can result in the optimizer making improvements locally without working to find the most performant overall structure. Another option for parameterization is to use B-splines as the basis functions for the level set \cite{Nol2020,Wang2019,Nol2022}. These approaches increase the level of smoothness of the level set without having to use high order \acp{fe}. However, although the increased smoothness eliminates some of the need for filtering techniques, the design parameters are still only capable of influencing a local region of the domain. Methods that allow for a larger influence of design parameters on the \ac{ls} function have also been proposed \cite{Pingen2009,Liu2020,Wang2006}. These methods use large supports for the basis functions by utilizing, for example, radial basis functions to allow for wider influence of a decreased number of design parameters. Using only parameters which have a widespread influence can however result in an inability to describe small spatial variations of the interface. 

To incorporate design parameters that simultaneously optimize the geometry at multiple scales and retain the advantages of both local and widespread design parameter influence, we propose to parameterize the \ac{ls} using a modified U-net convolutional neural network. Here, the design variables have an influence at a scale that depends on the layer that they appear within the network. Using this method, we can discover geometries that have greater regularity than those using only local design parameters by making improvements to large-scale features during evolution. We also, however, maintain the ability to describe small spatial variations of the interface. The nonlinearity of the U-net and the nature of the connectivity of the parameters allow for the discovery of complex relationships between features at different scales which can ultimately lead to the emergence of high-performing regular geometries and avoid the sub-optimal solutions of common parameterizations \cite{Yan2018}. It is noted that some regularity can be added to the previously discussed methods, but it often requires perturbation of the objective function, see e.g. \cite{Dijk2012}.} 

\begin{comment}
As an alternative, we introduce a
neural parameterization of the geometry. We set our design variables in
this case to be the parameters of a particular artificial \ac{nn} that outputs the \ac{ls} function  \ac{DOF}. Performing this step, we obtain control over the optimization
problem by controlling the connectivity of parameters, allowing them to influence cells spread across the domain. Although the expressivity is unchanged, since the \ac{ls}s are in the same space, the parameters controlling the evolution act to optimize the geometry at multiple scales. The ultimate objective is then for the optimization process to unveil regularized geometries with good performance. 
\end{comment}

The combination of machine learning and \ac{to} was explored as early as the 1990s \cite{Adeli1995} but has gained massive momentum in recent years \cite{Zhang2021,Woldseth2022}. \acp{nn} and other ML techniques can be incorporated into the \ac{to} process in many ways.
Common data-driven approaches attempt to train networks to map problem descriptions directly to a geometry \cite{Hoang2022,Yu2018,Li2019,Zheng2021}. These however require pre-training on already optimized samples and suffer from a lack of generalisability \cite{Woldseth2022}. Others replace some or all of the optimization loop for accelerated convergence by training an auxiliary network \cite{Kallioras2020,Joo2021}. These approaches are based on the premise, which in general is not necessarily true, that early iterations of the optimization contain the information to produce performant optimal geometries. An alternative method is the inclusion of a \ac{nn} for a parameterization of the geometry \cite{Deng2020,Chandrasekhar2020,Hoyer2019}. These approaches typically optimize the parameters of a \ac{nn} representing a continuous function that maps positions in space to a density. These approaches tend to focus on reducing the dimensionality of the design space assuming that \acp{nn} can efficiently achieve expressiveness with a small number of parameters \cite{Barron1994}. A reduction in parameters does not, however, necessarily lead to faster convergence for \acp{nn} \cite{Chandrasekhar2020} compared to the standard SIMP approach. 

Instead of focusing on a neural parameterization that reduces the dimensionality of the problem, we select a network description of the geometry which is specifically designed to learn effectively on problems involving the segmentation of a domain. The network used in this case is a modification of the U-Net convolutional network. The U-Net architecture was originally developed for biomedical
image segmentation tasks \cite{ronneberger2015unet} but has proven
successful for a variety of applications in which multi-scale features
and spatial correlation is important \cite{Ulyanov2020}. These networks are typically composed of encoding and decoding halves. The encoding section maps the context of input images into a low dimensional latent space which is localized in the upsampling section to provide segmentation at the desired resolution. The work in \cite{Hoyer2019} exploited the properties of this network showing
improved performance with a U-Net density parameterization for a SIMP
structural optimization problem. Similar to \cite{Hoyer2019}, we
use a trainable input vector for the network and feed this into the
up-sampling half of the U-Net. In contrast to most applications of this
network, we have no input image and therefore do not need the
encoding of half of the network. It is the up-sampling (or decoding) part of
the network that provides the parameterization of the
multi-scale features which are important in this context. 

\sbcom{
When using \ac{ls} \ac{to} methods, the optimized designs can be dependent on the initialization of the design parameters. This is because sensitivity information exists only on the design boundary making it difficult to discover where the introduction of a new hole away from the boundary in the design domain would be beneficial. Depending on the application, different approaches to initializing the domain can be made. For small-scale feature design, one may choose to seed the geometry with irregular structure as in \cite{Dapogny2019} but for most level set methods, evenly spaced holes are usually the starting point \cite{Jenkins2016,Villanueva2017,Kreissl2012}. The final design in these applications is then dependent on the size, shape and location of the holes \cite{Barrera2020}. In our approach, we alleviate some of the dependency of initialization by taking many random seeds for the starting geometry. 
}

\sbcom{
Using unfitted/immersed techniques is currently the leading way to retain accurate descriptions of the interface in topology optimization \cite{Alexandersen2020}. These methods are typically computationally expensive, however, because finite difference schemes are required for the computation of terms in the backward pass \cite{Sharma2016}. To the best of our knowledge, we are the first to implement a fully automatically differentiable framework for an embedded method in topology optimization so that the backward pass takes roughly the same time as solving the forward problem.
}

The Julia \cite{Julia-2017} programming language is used to implement all aspects of this project with the \ac{fe} toolbox Gridap \cite{Badia2020,Verdugo2022} being the main package utilized. We also use the Julia machine learning library Flux \cite{Flux.jl-2018} for the implementation of the U-Net. Using these foundations, we implement a routine combining \acp{nn} with an unfitted \ac{fe} based \ac{to}. The main contributions of this work are the presentation of:
\begin{itemize}
	\item An unfitted \ac{ls} \ac{to} method with a \ac{nn} parameterization that avoids sub-optimal solutions and achieves regular optimized geometries without handcrafted initializations and 
	
	\item a fully automatically differentiable unfitted \ac{ls} \ac{to} method for multiphysics problems with complex boundary conditions.
\end{itemize}
We present the overall framework as follows. First, in Section \ref{optimsation-problem}, we present the entire
optimization loop at a high level. We then go into more detail about
various stages in the loop. Details of the architecture of the neural
network are found in Section \ref{neural-architecture}, details of the geometry processing are presented in Section \ref{level-set-function-processing}, the numerical discretization of the problem is presented in Section \ref{simulation} and the gradient implementation in Section \ref{backwards-pass-implementation}. We then benchmark the method against baseline methods and show the generality of the method with an application to a multiphysics problem with complex boundary conditions in Section \ref{numerical-experiments}.

%% file: inverseproblem.tex
\section{optimization Problem}\label{optimsation-problem}

%\the\linewidth

%\begin{document}
In this section, we provide a succinct overview of the overall \ac{to} algorithm proposed in this work. 
We aim to solve the problem:
	\begin{equation}
		\begin{aligned}
			\underset{\mathbf{p}}{\text{min}} & \ J(\boldsymbol{u}(\mathbf{p}),\mathbf{p})&  \\
			\text{s.t} & \ \mathscr{R}(\boldsymbol{u}(\mathbf{p}),\mathbf{p}) &= 0 ,\\
				   & \ \quad \quad \  \mathscr{V}(\mathbf{p})     &= 0 ,\\
		\end{aligned}
	\label{eq:opt-problem}
	\end{equation}

where 
$\mathbf{p}$ are the parameters that describe our geometry,
  $\mathscr{V}$  is an equality constraint (e.g. for the volume),
  $\mathscr{R}$  is the PDE residual,
  $\boldsymbol{u}$ is the solution of the PDE and
  $J$  is the objective.
If desired, further equality and inequality constraints can then be imposed by adding penalty terms to the objective function.

\begin{figure}
	\centering
    \def\svgwidth{0.55\linewidth}
    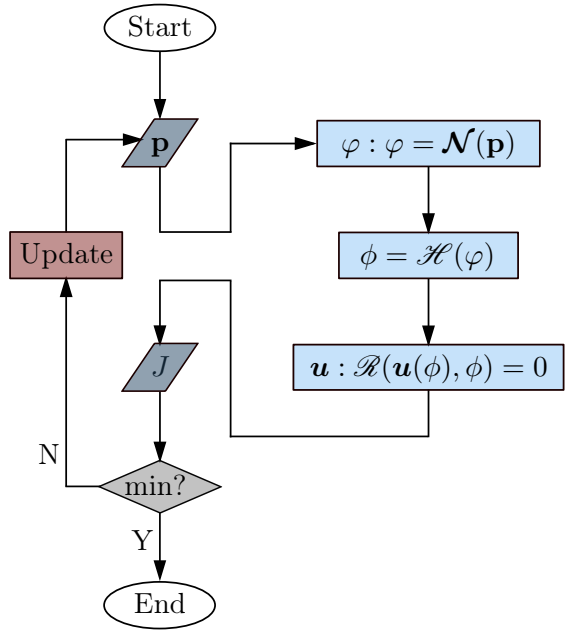

	\caption{Computational graph of the optimization loop. We start with an input to the system $\mathbf{p}$, and perform the forward pass by descending through the blue boxes on the right side to obtain a performance measure $J$ as explained in Section \ref{forward-pass}. %We evaluate the network to obtain $\mathbf{\varphi}$, apply the constraint to obtain $\phi$ and solve the residual equation to obtain $\boldsymbol{u}$.
 	The convergence criteria are used to decide whether this $J$ represents an acceptable minimum. If not, the backward pass is performed to compute an update for the parameters as explained in Section \ref{backwards-pass} and the loop is continued. %\sbcom{\hl{I would change the second box to $\phi = \mathscr{H}(\mathbf{\varphi})$, where $\mathscr{H}$ is an operator that takes the NN and generates a suitable  function (e.g., after re-distancing and volume correction by \emph{traslation}).}} } 
	}
 	\label{fig:FlowChart}
\end{figure}

\subsection{Optimization Loop}\label{optimization-loop}
To optimize the parameters $\mathbf{p}$, we make use of a gradient-based optimization strategy. To do so, we establish a map between $\mathbf{p}$ and $J$ and a means to compute the gradient $\frac{dJ}{d\mathbf{p}}$ for parameter updates.
For the neural \ac{ls} \ac{to} method, $\mathbf{p}$ represents the parameters of a particular \ac{nn} that outputs a vector $\mathbf{\varphi}$. This vector is processed using the operator $\mathscr{H}$ to obtain the \ac{ls} $\phi$ used in the PDE and objective function. %\sbcom{\hl{As I say in the figure, I would add some notation.}}. 
It is also only through $\phi$ that the PDE and objective depend on the parameter vector $\mathbf{p}$. 
With these definitions, we present the optimization loop for solving (\ref{eq:opt-problem}) in Figure \ref{fig:FlowChart}.

\subsubsection{Forward Pass}\label{forward-pass}

To solve the forward problem and get a performance measure $J$ for a set of parameters $\mathbf{p}$ we descend through the light blue boxes on the right-hand side of Figure \ref{fig:FlowChart} by performing the following steps:
 
\begin{enumerate}	
	\item In the first light blue box, we evaluate the network $\boldsymbol{\mathcal{N}}: \mathbf{p} \in \mathbb{R}^{N_p} \mapsto  \mathbf{\varphi} \in \mathbb{R}^{N}$, where $N_p$ is the number of parameters and $\boldsymbol{\mathcal{N}}$ is as defined in Section \ref{neural-architecture}. 
	\item 
	In the second light blue box, we process the output of the network $\mathbf{\varphi}$ using the operator $\mathscr{H}: \mathbf{\varphi} \in \mathbb{R}^{N} \mapsto \phi \in V_h^1$ to obtain a suitable \ac{ls} description of the geometry $\phi $, where $V_h^1$ is the \ac{fe} space for the \ac{ls} defined in Section \ref{level-set-function-processing}. This involves an interpolation on a \ac{fe} space, smoothing and the inclusion of an equality constraint on the geometry so that the final \ac{ls} function satisfies $\mathscr{V}(\phi)=0$. This step is broken down in Section \ref{level-set-function-processing}. We enforce the equality constraint here to allow for the use of an unconstrained optimization method suitable for \ac{nn}s.	
	\item In the third light blue box, we solve the \ac{fe} problem associated with the weak form of the residual $\mathscr{R} (\boldsymbol{u}_h(\phi),\phi)=0$ on the domain segmentation defined by $\phi$ for an approximate solution $\boldsymbol{u}_h$ obtained using a \ac{fe} discretization. Details of the \ac{fcm} used to solve the problem are given in Section \ref{simulation}. We can then evaluate the objective $J(\boldsymbol{u}_h(\phi),\phi) \in \mathbb{R}$.

\end{enumerate}

\subsubsection{Backwards Pass}\label{backwards-pass}

To perform the update for the parameters using a steepest descent optimization strategy, we require an evaluation of the gradient $\frac{\partial J}{\partial \mathbf{p}}$. 
To do this efficiently at a cost roughly matching that of the forward pass, we use reverse mode differentiation and define rules to propagate sensitivities through each of the steps in the forward pass. We use an adjoint rule for the PDE and use automatic differentiation for all of the partial derivatives, including the derivative of integrals with respect to the \ac{ls}.
The derivative is then passed onto a chosen optimizer to update the parameters $\mathbf{p}$. The implementation of the gradient computation is discussed in Section \ref{backwards-pass-implementation}. \sbcom{Using this method, arbitrary loss functions can be used for the physical problem operating on the simulation output fields and update directions for the neural network parameters can be computed to improve the performance at each iteration with respect to the chosen loss function.}

%% file: figures/FlowChart3.pdf_tex
%% Creator: Inkscape inkscape 0.92.5, www.inkscape.org
%% Accompanies image file 'FlowChart3.pdf' (pdf, eps, ps)
\begingroup%
  \makeatletter%
  \providecommand\color[2][]{%
    \errmessage{(Inkscape) Color is used for the text in Inkscape, but the package 'color.sty' is not loaded}%
    \renewcommand\color[2][]{}%
  }%
  \providecommand\transparent[1]{%
    \errmessage{(Inkscape) Transparency is used (non-zero) for the text in Inkscape, but the package 'transparent.sty' is not loaded}%
    \renewcommand\transparent[1]{}%
  }%
  \providecommand\rotatebox[2]{#2}%
  \newcommand*\fsize{\dimexpr\f@size pt\relax}%
  \newcommand*\lineheight[1]{\fontsize{\fsize}{#1\fsize}\selectfont}%
  \ifx\svgwidth\undefined%
    \setlength{\unitlength}{216.15159403bp}%
    \ifx\svgscale\undefined%
      \relax%
    \else%
      \setlength{\unitlength}{\unitlength * \real{\svgscale}}%
    \fi%
  \else%
    \setlength{\unitlength}{\svgwidth}%
  \fi%
  \global\let\svgwidth\undefined%
  \global\let\svgscale\undefined%
  \makeatother%
  \begin{picture}(1,0.99591523)%
    \lineheight{1}%
    \setlength\tabcolsep{0pt}%
    \put(0,0){\includegraphics[width=\unitlength,page=1]{FlowChart3.pdf}}%
    \put(0.17901289,0.59758967){\color[rgb]{0,0,0}\makebox(0,0)[lt]{\lineheight{1.25}\smash{\begin{tabular}[t]{l}\makebox[0pt]{Update}\end{tabular}}}}%
    \put(0.31932002,0.94005104){\color[rgb]{0,0,0}\makebox(0,0)[lt]{\lineheight{1.25}\smash{\begin{tabular}[t]{l}\makebox[0pt]{Start}\end{tabular}}}}%
    \put(0,0){\includegraphics[width=\unitlength,page=2]{FlowChart3.pdf}}%
    \put(0.31919601,0.42789876){\color[rgb]{0,0,0}\makebox(0,0)[lt]{\lineheight{1.25}\smash{\begin{tabular}[t]{l}\makebox[0pt]{$J$}\end{tabular}}}}%
    \put(0.31225625,0.24811583){\color[rgb]{0,0,0}\makebox(0,0)[lt]{\lineheight{1.25}\smash{\begin{tabular}[t]{l}\makebox[0pt]{min?}\\\end{tabular}}}}%
    \put(0,0){\includegraphics[width=\unitlength,page=3]{FlowChart3.pdf}}%
    \put(0.31878606,0.07025742){\color[rgb]{0,0,0}\makebox(0,0)[lt]{\lineheight{1.25}\smash{\begin{tabular}[t]{l}\makebox[0pt]{End}\end{tabular}}}}%
    \put(0,0){\includegraphics[width=\unitlength,page=4]{FlowChart3.pdf}}%
    \put(0.31932002,0.7668155){\color[rgb]{0,0,0}\makebox(0,0)[lt]{\lineheight{1.25}\smash{\begin{tabular}[t]{l}\makebox[0pt]{$\mathbf{p}$}\end{tabular}}}}%
    \put(0,0){\includegraphics[width=\unitlength,page=5]{FlowChart3.pdf}}%
    \put(0.72606592,0.59644261){\color[rgb]{0,0,0}\makebox(0,0)[lt]{\lineheight{1.25}\smash{\begin{tabular}[t]{l}\makebox[0pt]{$\phi = \mathscr{H}(\varphi) $ } \end{tabular}}}}%
    \put(0,0){\includegraphics[width=\unitlength,page=6]{FlowChart3.pdf}}%
    \put(0.72074825,0.76498652){\color[rgb]{0,0,0}\makebox(0,0)[lt]{\lineheight{1.25}\smash{\begin{tabular}[t]{l}\makebox[0pt]{$ \mathbf{\varphi} : \mathbf{\varphi} = \boldsymbol{\mathcal{N}}(\mathbf{p})$}\end{tabular}}}}%
    \put(0.72074825,0.42789876){\color[rgb]{0,0,0}\makebox(0,0)[lt]{\lineheight{1.25}\smash{\begin{tabular}[t]{l}\makebox[0pt]{$\boldsymbol{u}:\mathscr{R}(\boldsymbol{u}(\phi),\phi)=0$}\end{tabular}}}}%
    \put(0.29263476,0.16529173){\color[rgb]{0,0,0}\makebox(0,0)[lt]{\lineheight{1.25}\smash{\begin{tabular}[t]{l}\makebox[0pt]{Y}\end{tabular}}}}%
    \put(0.15291698,0.29534151){\color[rgb]{0,0,0}\makebox(0,0)[lt]{\lineheight{1.25}\smash{\begin{tabular}[t]{l}\makebox[0pt]{N}\end{tabular}}}}%
  \end{picture}%
\endgroup%

%% file: network.tex
\section{Network Architecture}\label{neural-architecture}

\begin{figure}
	\centering
	\setlength{\abovecaptionskip}{-10pt plus 1pt minus 20pt }
    \def\svgwidth{1\linewidth}
    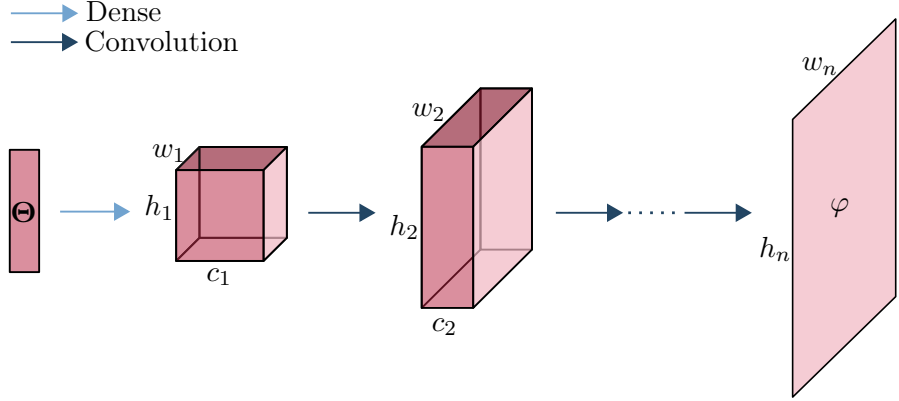
 
	\caption{Architecture of the \ac{nn}. A trainable input vector $\boldsymbol{\Theta}$ is fed into the network. The light blue arrow involves a set of operations that include a fully connected layer and the dark blue arrow involves a set of operations that include a convolutional layer. The intermediate data structures are of size $(c_l,w_l,h_l)$ and the final output, after $n$ layers, gives $\mathbf{\varphi}$. %\sbcom{\hl{I would make use of the width of the page, separating things further, so not arrows over boxes, and put the sizes right in the middle of edges.}}}
	}
	\label{fig:Network}
\end{figure}

In this section, we present the \ac{nn} architecture used in the geometry parameterization. This describes the mapping between the parameters $\mathbf{p}$ and the \ac{ls} vector $\mathbf{\varphi}$. \sbcom{The network used here is based on the U-net architecture and is mainly built from convolutional layers}. In contrast to a fully connected network, convolutional networks connect smaller sets of neurons in each layer assuming that neurons in close proximity have a more important relationship. This is a natural relaxation for the processing of spatial data. Furthermore, convolutional networks are made efficient by the assumption that features that are found in one local block are likely to be found in a different local block, i.e. somewhere else in the domain. This is done by sharing parameters amongst local blocks in the form of a convolutional filter. 
The specific architecture used in this work is the one presented in \cite{Hoyer2019}. Because we are simply reparameterizing the pixel values, there is no input into the network in the traditional sense. Most approaches combining \ac{nn}s and \ac{to} use fully connected layers and take the network to provide a map between a spatial input $\mathbf{x}$ and a scalar output $f_\theta (\mathbf{x})$ \cite{Deng2020,Chandrasekhar2020}. In our case, we only need a single evaluation of the network to output a vector that represents the entire discrete \ac{ls} function. \sbcom{ This is because, as in \cite{Hoyer2019}, the dimension of the output of the network matches the number of nodal values of the \ac{ls} function so that we have a 1:1 mapping between the network and the geometry discretization. This follows the typical use of the U-net, where the output image gives a segmented domain \cite{ronneberger2015unet}. In contrast, however, to original applications of the U-net, our input vector is also taken to be a set of trainable parameters $\mathbf{\boldsymbol{\Theta}}$.}

The architecture of the network is illustrated in Figure \ref{fig:Network}. The first arrow indicated with the label Dense in Figure \ref{fig:Network} contains a fully connected layer and a reshape:
\begin{equation}
	\mathbf{x}^{(1)}(\boldsymbol{\Theta}) = \mathrm{reshape}  (  \mathbf{W}  \boldsymbol{\Theta} + \mathbf{b}  ), \quad  \mathbf{x}^1 \in \mathbb{R}^{c_1,w_1,h_1},
\end{equation}
where $\mathbf{W} \in \mathbb{R}^{c_1 w_1 h_1, N_{\boldsymbol{\Theta}}}$ is the dense weight matrix, $\mathbf{b} \in \mathbb{R}^{c_1 w_1 h_1 }$ is the bias and $\mathrm{reshape}: \mathbb{R}^{c_1 w_1 h_1} \rightarrow \mathbb{R}^{c_1,w_1,h_1}$ is a reshaping map. The lengths $N_{\boldsymbol{\Theta}}$, $c_1$, $w_1$ and $h_1$ represent the length of $\boldsymbol{\Theta}$, the initial number of channels, the latent space image width and the latent space image height, respectively. Note here that the first nonlinearity is imposed at the beginning of the next layer.

The upsampling convolutional layers, depicted by the dark blue arrow in Figure \ref{fig:Network}, are defined as: 
\begin{equation}
	\mathbf{x}^{(l+1)}(\mathbf{x}^{(l)}) =  \boldsymbol{\mathcal{P}}^{(l)} (  \boldsymbol{\varrho}  ( \mathbf{\Phi}^{(l)}  (  \text{tanh}  ( \mathbf{x}^{(l)} )))), \quad  \mathbf{x}^{(l)} \in \mathbb{R}^{c_l  ,  w_l  ,  h_l}, \ \  \mathbf{x}^{(l+1)} \in \mathbb{R}^{c_{l+1}  ,  w_{l+1}  ,  h_{l+1}}, 
\end{equation}
where
$\mathbf{\Phi}^{(l)}: \mathbb{R}^{ c_l, w_l, h_l } \rightarrow \mathbb{R}^{ c_l, w_{l+1}, h_{l+1} }$ is a bilinear resize, 
$\boldsymbol{\varrho}$ is a
normalization to a mean of $0$ and variance of $1$ across the channel dimension and 
$\boldsymbol{\mathcal{P}}^{(l)}: \mathbb{R}^{ c_l  ,  w_{l+1}  ,  h_{l+1}} \rightarrow \mathbb{R}^{ c_{l+1}  ,  w_{l+1}  ,  h_{l+1} }$ is a convolutional operator with kernel size $(5,5)$ \sbcom{which is found to allow for the required level of expressivity.}

In this approach, our input to the network $\boldsymbol{\Theta}$ is taken to be trainable. So to define our parameter vector $\mathbf{p}$, we collect the parameters of $\mathbf{W}$, $\mathbf{b}$, $\boldsymbol{\Theta}$ and $\boldsymbol{P}^i$ into a vector $\mathbf{p}\in \mathbb{R}^{N_p} $. Then, by composing the layers, we obtain the function 
$\boldsymbol{\mathcal{N}}: \mathbf{p} \in \mathbb{R}^{N_p} \mapsto  \mathbf{\varphi} \in \mathbb{R}^{N}$:
\begin{equation}	
	{\boldsymbol{\mathcal{N}}} = \mathbf{x}^{(n)}(... ( \mathbf{x}^{(2)}(\mathbf{x}^{(1)} (\boldsymbol{\Theta})))).
\end{equation}
\sbcom{The output image of size $(w_n,h_n)$ obtained by evaluating the network at a set of parameters $\mathbf{\varphi}=\boldsymbol{\mathcal{N}}(\mathbf{p})$ is then used to define the geometry for the problem as described in Section \ref{level-set-function-processing}.
} Importantly, the locality is preserved when defining this function. 
As we traverse the network, we follow the design principle of the upsampling section of the U-Net and trade-off channel depth for spatial resolution. This means that, in general, the widths $w$ and heights $h$ will increase as we move through the network and the number of channels $c$ will decrease as we move through the network. The exact trade-offs here can vary, but we must set the resizes and channel refinement to ensure $h_n w_n = N $ and $ c_n = 1$ so that our output $\mathbf{\varphi}$ makes sense as a vector representing the  \ac{DOF} values for the \ac{ls}. 
We can increase the network size in the dense layer by increasing $N_{\boldsymbol{\Theta}}$, $w_1$ and $h_1$ and in the convolutional layers by increasing the elements in $c$ and the number of layers. 

\sbcom{
One approach to initializing the parameters of the \ac{nn} is to pre-train the network to output the manually selected geometry with holes, as in \cite{Deng2021}. Using this as an initial guess, however, causes the geometry to converge quickly to poor local minima. A more common approach when working with \ac{nn}s and a given objective map is to start with small random weights \cite{RUMELHART1988}. In this case, we have high asymmetry in the weights and little activation function saturation. It turns out that a random initialization of the \ac{nn} with the volume constraint gives an initial guess of a domain with a few holes in random locations. This is in contrast to initializing the \ac{ls} \ac{fe} function \ac{DOF} values with random values which gives a geometry with many small holes and fine features which is not necessarily desirable \cite{Barrera2020}. As is common in \ac{nn} approaches, we can then easily take multiple seeds for the geometry using different random initializations to alleviate initialization dependency. The random initializations simply correspond to different size holes in different locations. Using this method we eliminate the need for manual geometry initialization.

We use Xavier uniform initialization for the weights \cite{Glorot2010UnderstandingTD}, in this case the design variables, of the neural network to facilitate convergence. To minimize the effect of the initialization dependency of the optimized geometry, that is, wherever the holes appear in the initial design based on the particular random initialization, we take 100 random seeds of the parameters, run the optimization and take the final design with the lowest objective function value. 
}

%% file: figures/Network.pdf_tex
%% Creator: Inkscape inkscape 0.92.5, www.inkscape.org
%% Accompanies image file 'Network.pdf' (pdf, eps, ps)
\begingroup%
  \makeatletter%
  \providecommand\color[2][]{%
    \errmessage{(Inkscape) Color is used for the text in Inkscape, but the package 'color.sty' is not loaded}%
    \renewcommand\color[2][]{}%
  }%
  \providecommand\transparent[1]{%
    \errmessage{(Inkscape) Transparency is used (non-zero) for the text in Inkscape, but the package 'transparent.sty' is not loaded}%
    \renewcommand\transparent[1]{}%
  }%
  \providecommand\rotatebox[2]{#2}%
  \newcommand*\fsize{\dimexpr\f@size pt\relax}%
  \newcommand*\lineheight[1]{\fontsize{\fsize}{#1\fsize}\selectfont}%
  \ifx\svgwidth\undefined%
    \setlength{\unitlength}{453.54330709bp}%
    \ifx\svgscale\undefined%
      \relax%
    \else%
      \setlength{\unitlength}{\unitlength * \real{\svgscale}}%
    \fi%
  \else%
    \setlength{\unitlength}{\svgwidth}%
  \fi%
  \global\let\svgwidth\undefined%
  \global\let\svgscale\undefined%
  \makeatother%
  \begin{picture}(1,0.34375)%
    \lineheight{1}%
    \setlength\tabcolsep{0pt}%
    \put(0.1972604,0.33237484){\color[rgb]{0,0,0}\makebox(0,0)[lt]{\begin{minipage}{0.10508791\unitlength}\raggedright Dense\end{minipage}}}%
    \put(0.19623986,0.30760233){\color[rgb]{0,0,0}\makebox(0,0)[lt]{\begin{minipage}{0.17116704\unitlength}\raggedright Convolution\end{minipage}}}%
    \put(0,0){\includegraphics[width=\unitlength,page=1]{Network.pdf}}%
    \put(0.13514959,0.14986327){\color[rgb]{0,0,0}\makebox(0,0)[lt]{\lineheight{1.25}\smash{\begin{tabular}[t]{l}$\boldsymbol{\Theta}$\end{tabular}}}}%
    \put(0,0){\includegraphics[width=\unitlength,page=2]{Network.pdf}}%
    \put(0.78611474,0.27513066){\color[rgb]{0,0,0}\makebox(0,0)[lt]{\lineheight{1.25}\smash{\begin{tabular}[t]{l}$w_n$\end{tabular}}}}%
    \put(0,0){\includegraphics[width=\unitlength,page=3]{Network.pdf}}%
    \put(0.80943928,0.1585463){\color[rgb]{0,0,0}\makebox(0,0)[lt]{\lineheight{1.25}\smash{\begin{tabular}[t]{l}$\mathbf{\varphi}$\end{tabular}}}}%
    \put(0.7510654,0.12069275){\color[rgb]{0,0,0}\makebox(0,0)[lt]{\lineheight{1.25}\smash{\begin{tabular}[t]{l}$h_n$\end{tabular}}}}%
    \put(0,0){\includegraphics[width=\unitlength,page=4]{Network.pdf}}%
    \put(0.24555259,0.15487569){\color[rgb]{0,0,0}\makebox(0,0)[lt]{\lineheight{1.25}\smash{\begin{tabular}[t]{l}$h_1$\end{tabular}}}}%
    \put(0.29659512,0.10057073){\color[rgb]{0,0,0}\makebox(0,0)[lt]{\lineheight{1.25}\smash{\begin{tabular}[t]{l}$c_1$\end{tabular}}}}%
    \put(0,0){\includegraphics[width=\unitlength,page=5]{Network.pdf}}%
    \put(0.2518225,0.20308445){\color[rgb]{0,0,0}\makebox(0,0)[lt]{\lineheight{1.25}\smash{\begin{tabular}[t]{l}$w_1$\end{tabular}}}}%
    \put(0.44707518,0.13898735){\color[rgb]{0,0,0}\makebox(0,0)[lt]{\lineheight{1.25}\smash{\begin{tabular}[t]{l}$h_2$\end{tabular}}}}%
    \put(0.48166175,0.05927564){\color[rgb]{0,0,0}\makebox(0,0)[lt]{\lineheight{1.25}\smash{\begin{tabular}[t]{l}$c_2$\end{tabular}}}}%
    \put(0,0){\includegraphics[width=\unitlength,page=6]{Network.pdf}}%
    \put(0.46539154,0.23692574){\color[rgb]{0,0,0}\makebox(0,0)[lt]{\lineheight{1.25}\smash{\begin{tabular}[t]{l}$w_2$\end{tabular}}}}%
    \put(0,0){\includegraphics[width=\unitlength,page=7]{Network.pdf}}%
  \end{picture}%
\endgroup%

%% file: lsprocessing.tex
\section{\ac{ls} Function Processing}\label{level-set-function-processing}

In this section, we detail the \ac{ls} description of the geometry used in the method. We explain how the output vector from the \ac{nn} $\mathbf{\varphi}$ is used to define the \ac{ls} to be used in the numerical method. The computation of the \ac{ls} function is performed in four steps. We introduce an interpolation step in Section \ref{interpolation} to obtain a first \ac{ls} function $\phi_{n(1)}$. In Section \ref{smoothing}, we smooth out this \ac{ls} to obtain a smooth \ac{ls} function $\phi_{f(2)}$. Then, we perform a reinitialization of that \ac{ls} in Section \ref{reinitialization} to obtain an \ac{ls} function $\phi_{s(3)}$. Finally, in Section \ref{translation}, we propose a volume correction strategy to end up with the final \ac{ls} function $\phi_{b(4)}$. In the subsequent sections, we will refer to this final \ac{ls} function $\phi_{b(4)}$ as $\phi$.

\subsection{Interpolation}\label{interpolation}

 In this method, we work with discrete \ac{ls} functions $\phi \in V_h^1$. The  \ac{DOF} of $V_h^1$ are the values of the function at the vertices of $\mathcal{T}_h$, which we denote with $\{\boldsymbol{x}_i\}_{i=1}^N$, where $N$ is the number of mesh nodes. Thus, there is an isomorphism between $V_h^1$ and $\mathbb{R}^N$. The output image vector of the \ac{nn} $\varphi$ are the \ac{DOF} values that uniquely determine the \ac{ls} \ac{fe} function.

\subsection{Smoothing}\label{smoothing}
Next, we convolve the function with a linear filter for smoothing:
\begin{equation}
  {\phi_{f(2)}}_i = (\sum^N_{j=1} w_{ij} )^{-1} ( \sum^N_{j=1} w_{ij} {\phi_{n(1)}}_j),
\end{equation}
where $w_{ij}=\max(0,r_f-\vert x_i-x_j \vert )$,  ${\phi_{n(1)}}_j$ is the $j^{th}$ degree of freedom of the function $\phi_{n(1)}$, ${\phi_{f(2)}}_i$ is the $i^{th}$ degree of freedom of the function $\phi_{f(2)}$ and $r_f$ is the smoothing radius.

\subsection{Reinitialization}\label{reinitialization}
We then reinitialize the \ac{ls} as a signed distance function. This is often done to gain control over the spatial gradient of the \ac{ls} function to improve convergence \cite{vanDijk2013}. In our case, it is of even greater importance as it also guarantees that when we apply the translation to the \ac{ls} to satisfy the volume constraint, we do not artificially introduce volumes into the domain far from the boundary where the \ac{ls} is close to zero. This would add discontinuity to the problem harming convergence. To perform this step, we solve the reinitialization equation in \cite{Xing2009} to obtain a signed distance function $\phi_{s(3)}\in V_h^1$:
\begin{equation}
  \frac{ \partial \phi_{s(3)} }{\partial \tau} + \text{sign}(\phi_{s(3)})(\vert \boldsymbol{\nabla}\phi_{s(3)} \vert - 1 ) = 0.
\end{equation}
The problem is solved using Picard iterations at steady state using the initial point $\phi_{f(2)}$ so that only one solve of the adjoint equation is required in the backward pass. Artificial viscosity is added to the problem for stabilization and a surface penalty term is integrated on the embedded boundary to prevent movement of the zero iso-surface in the reinitialization. The weak form of the problem is then to find the solution $\phi_{s(3)}\in V_h^1$ of the equation:
\begin{equation}
  \int_\Omega ( \boldsymbol{w}\cdot\boldsymbol{\nabla}(\phi_{s(3)})  v  + c_a h \vert \boldsymbol{w} \vert \boldsymbol{\nabla}(\phi_{s(3)}) \boldsymbol{\nabla} v  - \text{sign}(\phi_{s(3)}) v  ) d\Omega + \int_\Gamma (\phi_{s(3)}, v )d\Gamma = 0
\label{eq:Rs}
\end{equation}
for all $ v \in V_h^1$, where $\boldsymbol{w} = \text{sign}(\phi_{s(3)})(\boldsymbol{\nabla}(\phi_{s(3)})/ \vert\boldsymbol{\nabla}(\phi_{s(3)}) \vert )$, $c_a$ is the stabilization coefficient, set to $3$, and $h$ is the element size. 

\subsection{Translation}\label{translation}
The next task is to impose the volume constraint. This is done by applying a translation to the entire \ac{ls} function. Here we solve the nonlinear equation $\mathscr{V}$  for the scalar bias  $b$ to obtain a translation to the \ac{ls} function such that the domain $\Omega_{\rm in}$ satisfies the volume constraint. A bisection method is used to find the root $b$ of the equation:
\begin{equation}
  \mathscr{V} =  \int_{\Omega_{\rm in}} d\Omega - \mathscr{V}_0 = 0,
\end{equation}
where $\Omega_{\rm in}$ is defined by the \ac{ls} $\phi_{b(4)} = \phi_{s(3)} + b$ and $\mathscr{V}_0$ is the volume fraction given by the constraint. This is the \ac{ls} which is then used to define the boundary of the \ac{fe} problem. Details of how the \ac{ls} is used to define domains is given in the next Section.

%% file: FCM.tex
\section{Numerical Discretization}\label{simulation}

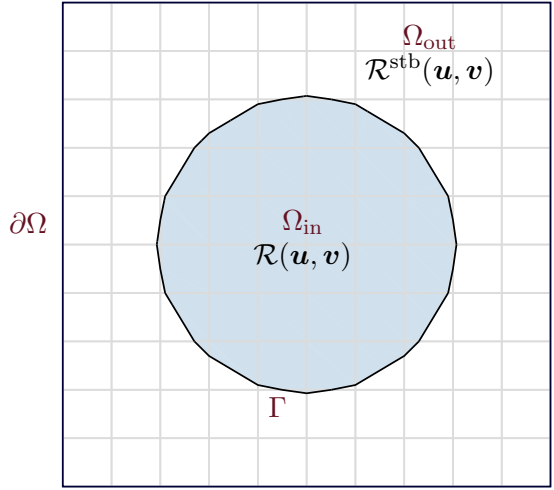
\begin{figure}
  \centering
    \def\svgwidth{0.55\linewidth}
    \import{./figures/}{DomainDefinition.pdf_tex}

  \caption{Illustration of the problem domains. The background domain is segmented into the domains $\Omega_{\rm in}$ and $\Omega_{\rm out}$. In $\Omega_{\rm in}$, the physical terms $\mathcal{R}(\boldsymbol{u},\boldsymbol{v})$ are integrated. In $\Omega_{\rm out}$, the stabilization terms $ \mathcal{R}^{\mathrm{stb}}(\boldsymbol{u},\boldsymbol{v})$ are integrated. }
  \label{fig:domain-definition}
\end{figure}

Here we formulate the numerical discretization of the PDEs being solved in the simulation step of the optimization loop. We consider a background polyhedral bounded domain $\Omega$ with boundary $\partial \Omega$. A \ac{ls} function $\phi(x)$ is used to split $\Omega$ into two subdomains $\Omega_{\rm in}(\phi)$ and $\Omega_{\rm out}(\phi)$ as follows (see Figure \ref{fig:domain-definition}):
\begin{equation}
  \begin{aligned}
     & \Omega_{\rm in}(\phi) = \{ x \in \Omega :  \phi(x) > 0 \},
    \quad \Omega_{\rm out}(\phi) = \{ x \in \Omega :  \phi(x) < 0 \}.
  \end{aligned}
  \label{eq:Omega_def}
\end{equation}
We denote the interface between these two subdomains as $\Gamma(\phi) \doteq \partial \Omega_{\mathrm{in}}(\phi) \cap \partial \Omega_{\mathrm{out}}(\phi)$.
Let us assume that $\partial \Omega_{\mathrm{in}}(\phi) \cap \partial \Omega \neq \emptyset$. The domain $\Omega_{\mathrm{in}}(\phi)$ is the one in which we consider our PDE problem. The weak form of the continuous problem can be stated as follows: find $\boldsymbol{u} \in V$ such that
\begin{equation}
  \int_{\Omega_{\mathrm{in}}(\phi)} \mathcal{R}(\boldsymbol{u},\boldsymbol{v}) \mathrm{d} \Omega = 0, \quad \forall v \in V.
  \label{eq:R}
\end{equation}
where $V$ is a Hilbert space in which the problem is well-posed. We consider zero flux Neumann boundary conditions on $\partial \Omega_{\mathrm{in}}(\phi) \setminus \partial \Omega$ and (for simplicity)  homogeneous boundary conditions on  $\partial\Omega_{\mathrm{in}}(\phi) \cap \partial \Omega$; the generalization to non-homogeneous Dirichlet boundary conditions is straightforward.

The domain $\Omega_{\mathrm{in}}(\phi)$ will change along the optimization process. As a result, it is not practical to compute body-fitted unstructured meshes for the geometrical discretization of $\Omega_{\mathrm{in}}(\phi)$. Instead, we consider a background mesh, which can simply be a Cartesian background mesh of $\Omega$  and make use of an unfitted \ac{fe} discretization. Unfitted (or embedded) discretizations relax the geometrical constraints but pose additional challenges to the numerical discretization \cite{Burman2010,Badia2018}. The first issue is the integration over cut cells (for details, see \cite{Badia2022-Geometrical}). The other issue is the so-called \emph{small cut cell} problem. Cut cells with arbitrary small support lead to ill-conditioned systems \cite{dePrenter2017}. Various techniques can be used to stabilize the problem including the \ac{cutfem} \cite{Burman2010}, the \ac{agfem} \cite{Badia2018} and the 
\ac{fcm} \cite{Parvizian2007}. 
We select the \ac{fcm} method in this case because it is differentiable with respect to the \ac{ls} 
everywhere in the domain. The \ac{cutfem} and \ac{agfem} are conversely not differentiable. %(see Proposition \ref{thm:CutFEM-nondiff}). %\sbcom{\hl{acronyms, I would use a package to be sure that we define it the first time we use them}}. The idea of \ac{fcm} method is to add a penalty term in the artificial domain $\Omega_{\mathrm{out}}$ that makes the discrete linear system solvable. Even though this method is not consistent and intrinsically low order, it is suitable for \ac{to} applications, which are intrinsically low-order.

In order to state the discrete form of the continuous problem, we introduce the unfitted \ac{fe} space. Let $\mathcal{T}_h$ represent a conforming, quasi-uniform and shape regular partition (mesh) of $\Omega$, $h$ being a characteristic mesh size. $\Omega$ can be a trivial geometry, e.g., a square or cube, and $\mathcal{T}_h$ can be a Cartesian mesh. We define a nodal Lagrangian \ac{fe} space of order $q \geq 1$ on $\mathcal{T}_h$ as:
\begin{equation}
  V_{h}^q = \{ \boldsymbol{v}_h \in \mathcal{C}^0(\Omega) : \boldsymbol{v}_h\vert_{K} \in \mathcal{X}_q(K) \ \forall K \in \mathcal{T}_h \},
\end{equation}
where $\mathcal{X}_q(K)$ is the space $\mathcal{Q}_q(K)$ of  polynomials with maximum degree $q$ for each variable when $\mathcal{T}_h$ is a quadrilateral or hexahedral mesh and the space $\mathcal{P}_q(K)$ of polynomials of total degree $q$ when $\mathcal{T}_h$ is a simplicial mesh. In this work, we consider low-order spaces, which is the most reasonable choice for \ac{to} applications.

The weak formulation of the problem solved by the method is now described. Let us represent with $V_{h,0}^q = V_h^q \cap \mathcal{C}_{0}^{0}(\bar{\Omega})$ the nodal \ac{fe} space that vanishes on the boundary $\partial \Omega$. 

Now, we can define a first-order \ac{fcm} discretization of (\ref{eq:R}) as follows: find $\boldsymbol{u}_h \in V_h^1$ such that
\begin{equation}
  \int_{\Omega_{\mathrm{in}}(\phi)} \mathcal{R}(\boldsymbol{u}_h,\boldsymbol{v}_h) \mathrm{d} \Omega +
  \int_{\Omega \setminus \Omega_{\mathrm{in}}(\phi)} \alpha_{\mathrm{out}} \mathcal{R}^{\mathrm{stb}}(\boldsymbol{u}_h,\boldsymbol{v}_h) \mathrm{d} \Omega = 0, \quad \forall v \in V_h^1,
\label{eq:RFCM}
\end{equation}
where $\alpha_{\mathrm{out}} \ll 1$ is the penalty parameter and $\mathcal{R}^{\mathrm{stb}}$ is a \emph{stabilizing}  differential operator on the artificial domain.

\newcommand{\ubs}{\boldsymbol{u}}

\subsection{Poisson Equation}\label{poisson-equations}

The poisson equation is used to model the temperature $\theta(\boldsymbol{x})$ 
that satisfies
\begin{equation}\protect\hypertarget{eq:T}{}{
    -\boldsymbol{\nabla}\cdot (\boldsymbol{\kappa} \boldsymbol{\nabla} \theta) = f \quad \text{in}\ \Omega_{\mathrm{in}}(\phi),
  }\label{eq:T}\end{equation}
where $\boldsymbol{\kappa}$  is the thermal conductivity tensor of the material and $f$ is a thermal source.
A zero Dirichlet condition ($\theta=0$) is prescribed on $\Omega_{\mathrm{in}}(\phi) \cap \partial \Omega$ and a zero flux condition ($\boldsymbol{n}\cdot(\boldsymbol{\kappa}\boldsymbol{\nabla}\theta) = 0$) is prescribed on $\Gamma(\phi)$. 

The \ac{fcm} approximation of this problem without a source term reads as: find $\theta_h \in V_{h,0}^{1}$ such that  
  \begin{equation}
  \int_{\Omega_{\mathrm{in}}(\phi)} \boldsymbol{\kappa} \boldsymbol{\nabla}\theta_h \cdot \boldsymbol{\nabla}v_h  \mathrm{d} \Omega +
  \int_{\Omega \setminus \Omega_{\mathrm{in}}(\phi)} \alpha_{\mathrm{out}} \boldsymbol{\kappa} \boldsymbol{\nabla}\theta_h \cdot \boldsymbol{\nabla}v_h \mathrm{d} \Omega = 0, \quad \forall v_h \in V_{h,0}^1.
  \label{eq:RT}
\end{equation}
One can readily check that this method is weakly enforcing the zero flux condition on $\Gamma(\phi)$ as $\alpha_{\mathrm{out}} \rightarrow 0$.  We observe that we use the same differential operator in the artificial domain for stabilization purposes (times the scaling coefficient $\alpha_{\mathrm{out}}$).

We consider the \ac{to} problem in which we aim at finding a level-set $\phi$ that minimizes the integral of the temperature: %\sbcom{\hl{name?}}
\begin{equation}
  J(\phi,\theta_h(\phi))  = \int_{\Omega(\phi)} \theta_h(\phi) \  \mathrm{d}\Omega,
\end{equation}
where $\theta_h(\phi)$ is the solution of (\ref{eq:RT}) given $\phi$.

\hypertarget{linear-elasticity}{%
  \subsection{Linear elasticity}\label{linear-elasticity}}

We want to obtain the displacement $\boldsymbol{d}(\boldsymbol{x})$ that satisfies the linear elasticity equation  
\begin{equation}\protect\hypertarget{eq:d}{}{
    \left.
    \begin{aligned}
      -\boldsymbol{\nabla}	\cdot \bm{\sigma} (\boldsymbol{d})  & = \boldsymbol{f} &  & \text{in} \   \Omega_{\rm in}(\phi),
    \end{aligned}
    \right.
  }\label{eq:d}\end{equation} where $ \bm{\sigma} = \lambda \text{tr}(\bm{\varepsilon})I+2\mu \bm{\varepsilon} $ is the stress tensor, $\bm{\varepsilon} = \frac{1}{2}(\boldsymbol{\nabla}\boldsymbol{d} + (\boldsymbol{\nabla}{\boldsymbol{d}})^\top)$ is the symmetric gradient, $ \(\boldsymbol{d}\)$ is the displacement, $\lambda$ and $\mu$ are the Lam\'e parameters given by $\lambda = (E\nu)/((1+\nu)(1-2\nu)) $ and $ \mu=E/(2(1+\nu)) $
and $\boldsymbol{f}$ is the forcing term. A zero Dirichlet condition ($\boldsymbol{d}=\boldsymbol{0}$) is prescribed on $\Omega_{\mathrm{in}}(\phi) \cap \partial \Omega$ and a zero stress condition ($\boldsymbol{n}\cdot \boldsymbol{\sigma}(\boldsymbol{d}) = 0$) is prescribed on $\Gamma(\phi)$.

The \ac{fcm} approximation of this problem without a forcing term reads as: find $\boldsymbol{d}_h \in \boldsymbol{V}_{h,0}^{1} \doteq  [V_{h,0}^1 ]^D$ such that  
  \begin{equation}
  \int_{\Omega_{\mathrm{in}}(\phi)} \boldsymbol{\sigma}(\boldsymbol{d}_h) : \boldsymbol{\varepsilon}(\boldsymbol{v}_h)  \mathrm{d} \Omega +
  \int_{\Omega \setminus \Omega_{\mathrm{in}}(\phi)} \alpha_{\mathrm{out}} \boldsymbol{\sigma}(\boldsymbol{d}_h) : \boldsymbol{\varepsilon}(\boldsymbol{v}_h)  \mathrm{d} \Omega = 0, \quad \forall \boldsymbol{v}_h \in \boldsymbol{V}_{h,0}^1.
  \label{eq:Rd}
\end{equation}
It is easy to check that the zero-stress condition on $\Gamma(\phi)$ is recovered as $\alpha_{\mathrm{out}} \rightarrow 0$. For the linear elasticity equation, we again use the same differential operator in the artificial domain for stabilization purposes.

A typical \ac{to} problem in solid mechanics is the minimization of the strain energy. In this case, we aim at finding a level-set $\phi$ that minimizes the cost function 
\begin{equation}
  J(\phi,\boldsymbol{d}_h(\phi))  = \int_{\Omega(\phi)} \bm{\sigma}(\boldsymbol{d}_h):\bm{\varepsilon}(\boldsymbol{d}_h) \  \mathrm{d}\Omega,
  \label{eq:Jd}
\end{equation}
where $\boldsymbol{d}_h(\phi)$ is the solution of (\ref{eq:Rd}) given $\phi$.

\subsection{Linear Elasticity with Fluid Forcing Terms}\label{FSI}

Once again, we want to obtain the displacement $\boldsymbol{d}_h(\boldsymbol{x})$ that satisfies the linear elasticity formulation in (\ref{eq:d}). In this case, however, we consider the surface traction exerted by the fluid:
\begin{equation}\protect\hypertarget{eq:rdGammas}{}{
  \begin{aligned}
    \int_{\Gamma(\phi)}
      ( \boldsymbol{n} \cdot \boldsymbol{\nabla}\boldsymbol{u}_h  - {p}_h \ \boldsymbol{n} ) \cdot \boldsymbol{v}
    {\rm d}x,
  \end{aligned}
}\label{eq:rdGammas}\end{equation}
where the fluid velocity $\boldsymbol{u}_h$ and pressure field ${p}_h$ are obtained by solving a fluid problem in the domain $\Omega_{\rm out}$. 
These fields are obtained by solving the Stokes equations with a Brinkmann penalization as in \cite{Borrvall2002} but without intermediate interpolation of permeabilities at the boundary.

In order to approximate the fluid problem, we use a mixed \ac{fe} method, namely the equal order pair $\boldsymbol{V}_{h,0}^{1} \times {V}_h^1$.

We find \((\boldsymbol{u}_h,{p}_h)\in \boldsymbol{V}_{h,0}^1 \times V_{h,0}^1 \)
such that:
\begin{equation}\protect\hypertarget{eq:rupOmega}{}{
  \begin{aligned}
    %\mathscr{R}_{\boldsymbol{u},p}^\Omega \doteq
    \int_{\Omega}%\cup\Omega_s\cup\Omega_f}
    &\left[
      \alpha \boldsymbol{u}_h \cdot \boldsymbol{\psi}_h +
      \mu \boldsymbol{\nabla}\boldsymbol{u}_h \cdot\boldsymbol{\nabla}\boldsymbol{\psi}_h 
      -{p}_h (\boldsymbol{\nabla}\cdot \boldsymbol{\psi}_h)
      -(\boldsymbol{\nabla}\cdot \boldsymbol{u}_h) {q}_h
      - h^2 \boldsymbol{\nabla}p_h \cdot \boldsymbol{\nabla}{q_h}
      \right] 
    {\rm d}x
    =0, \quad 
    \\
    &\forall \boldsymbol{\psi}_h,{q}_h \in \boldsymbol{V}_{h,0}^{1} \times {V}_h^1,  %, where $\mathscr{R}_{\boldsymbol{u},p}^\Omega$ is given by:
  \end{aligned}
}\label{eq:rupOmega}\end{equation}
where
\begin{equation}
  \left\lbrace
  \begin{aligned}
    \alpha & = 0        &  & \text{in } \Omega_{\rm out} \\%\cup \Omega_f,  \\
    \alpha & = \alpha_u &  & \text{in } \Omega_{\rm in} \\%\cup \Omega_s. \\
  \end{aligned}
  \right.
\end{equation}
using an artificial porosity $\alpha_u$ to make the fluid problem well-posed in the solid domain $\Omega_{\rm in}$ and enforce the no-slip boundary condition. In the fluid domain $\Omega_{\rm out}$ we recover the Stokes equations.

The \ac{to} problem once again involves finding a level-set $\phi$ that minimizes the elastic strain using (\ref{eq:Jd}). 

\subsection{Differentiability of the unfitted \ac{fe} solver}\label{sec:}

An important property for the convergence of a \ac{to} strategy is the notion of shape differentiability of the cost function. 
A functional under a PDE constraint is considered shape differentiable if the mapping $\phi \rightarrow J(\boldsymbol{u_h}(\phi),\phi)$ 
is differentiable at the admissible set of domains in $\Omega$ defined by $\phi$.
In this section, we discuss how the choice of \ac{fe} stabilization can affect this property. 

The model problem in \eqref{eq:RT} with a \ac{fcm} stabilization is equivalent to that of a typical two-phase conductivity problem. 
With a solution $\boldsymbol{u}_h\in H^1(\Omega)$, the functional $J(\boldsymbol{u_h},\phi)$ for this problem can be proven to be shape differentiable, see \cite[Theorem 4.9]{Allaire2021}. 

Conversely, unfitted techniques involving stabilization only in the vicinity of the boundary are in general not shape differentiable. 
Regions of non-differentiability arise (typically when the boundary crosses over mesh nodes) harming the convergence of the geometry to optimized solutions \cite{Sharma2016}. To see why this is the case, we investigate shape perturbations under the \ac{cutfem} and \ac{agfem} formulations.

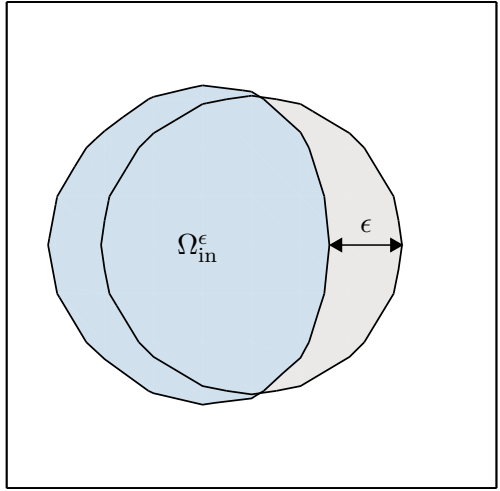
\begin{figure}
  \centering
    \def\svgwidth{0.55\linewidth}
    \import{./figures/}{Pertubation.pdf_tex}

  \caption{A small perturbation with size $\epsilon$ to the \ac{ls} to form a new domain $\Omega_{\rm in}^\epsilon$. }
  \label{fig:pertubation}
\end{figure}

\begin{figure}
  \centering
    \def\svgwidth{0.55\linewidth}
    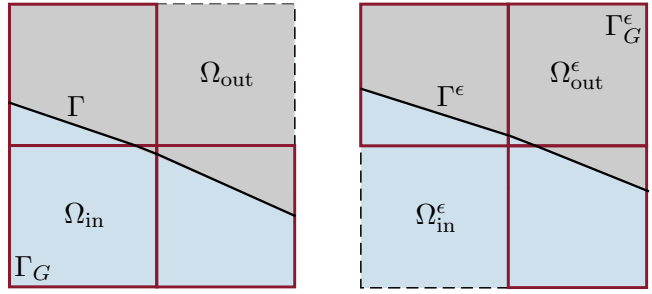

  \caption{The change in a portion of the ghost skeleton triangulation $\Gamma_G$, depicted by the red faces, before and after a perturbation to the boundary.}
  \label{fig:ghost-skeleton}
\end{figure}

If we were to use a restricted space for the solution and add ghost penalty terms in the vicinity of the interface following the CutFEM method \cite{CutFEM2015}, the problem is to find $u_h\in W_{h,0}^1.$ such that:
\begin{equation}
  \begin{aligned}
    \int_{\Omega_{\mathrm{in}}(\phi)} \mathcal{R}(\boldsymbol{u}_h,\boldsymbol{v}_h) \mathrm{d} \Omega_{\Omega_{\rm in}} +
    j(\boldsymbol{u}_h,\boldsymbol{v}_h)_{\Gamma_G} = 0, \quad \forall v_h \in W_{h,0}^1,
  \end{aligned}
\end{equation}
for a ghost penalty term $j$ on a ghost skeleton triangulation $\Gamma_G$ using the space $W_{h,0}^1$ as in \cite{CutFEM2015}. 
Consider the change to the domain $\Omega_{\rm in}$ from Figure \ref{fig:domain-definition} caused by a perturbation $\epsilon \delta$, where $\epsilon \in \mathbb{R}$ and $\delta\in V^1_h$, to the \ac{ls} function $\phi$. The resulting domain $\Omega_{\rm in}^\epsilon$ may be as in Figure \ref{fig:pertubation}. The ghost penalty term in this formulation does not depend on $\epsilon$ and instead changes depending on which cells are cut. Specifically, if $\Gamma^\epsilon$ crosses over a mesh node as in Figure \ref{fig:ghost-skeleton}, the ghost skeleton triangulation includes the faces of a new element. Non-zero terms are integrated on this triangulation introducing discontinuity to the problem with respect to the shape, since:
\begin{equation}
  \lim_{\epsilon \to 0} \  [ j(\boldsymbol{u}_h,\boldsymbol{v}_h)_{\Gamma_G^\epsilon} - j(\boldsymbol{u}_h,\boldsymbol{v}_h)_{\Gamma_G} ] \  \neq 0.
\end{equation}
With different terms added to the linear system which do not go to zero with $\epsilon$, we can see that the derivative of the solutions with respect to the shape can be ill-defined. A cost function operating on the solution could not, in general, be shape differentiable at these points.

In the \ac{agfem}, the problem is to find $u_h\in V^{agg}_{h,0}$ such that:
\begin{equation}
  \begin{aligned}
	  \mathscr{R}(\mathcal{E}(\boldsymbol{u}_h),\mathcal{E}(\boldsymbol{v}_h))_{\Omega_{\rm in}} \quad \forall v_h \in V^{agg}_{h,0},
  \end{aligned}
\end{equation}
for the extension operator $\mathcal{E}$ and space $V^{agg}_{h,0}$ as defined in \cite{Badia2018}. Similar to the CutFEM, regions of non-differentiability in the problem exist when the zero iso-surface of the \ac{ls} crosses over mesh nodes:
\begin{equation}
  \begin{aligned}
    \lim_{\epsilon \to 0} \ 
    [
      \mathscr{R}(\mathcal{E}(\boldsymbol{u}_h),\mathcal{E}(\boldsymbol{v}_h))_{\Omega_{\rm in}} 
    -
    \mathscr{R}(\mathcal{E}^\epsilon(\boldsymbol{u}_h),\mathcal{E}(\boldsymbol{v}_h))_{\Omega_{\rm in}^\epsilon} 
    ]
    \neq 0,
  \end{aligned}
\end{equation}
since $\mathcal{E}^\epsilon(\boldsymbol{u}) \neq \mathcal{E}(\boldsymbol{u}_h)$ in general because the support for cut cells potentially changes depending on which cells are cut. Following the same reasoning as above, the method is therefore not shape differentiable. The same is true for other methods which use stabilization approaches that act only in the vicinity of cut cells, e.g. \cite{Lang2014}, using similar branching strategies when crossing a node or reaching a certain threshold.
\qed
%\end{proof}

%% file: figures/domaindefinition.pdf_tex
%% Creator: Inkscape inkscape 0.92.5, www.inkscape.org
%% Accompanies image file 'domaindefinition.pdf' (pdf, eps, ps)
\begingroup%
  \makeatletter%
  \providecommand\color[2][]{%
    \errmessage{(Inkscape) Color is used for the text in Inkscape, but the package 'color.sty' is not loaded}%
    \renewcommand\color[2][]{}%
  }%
  \providecommand\transparent[1]{%
    \errmessage{(Inkscape) Transparency is used (non-zero) for the text in Inkscape, but the package 'transparent.sty' is not loaded}%
    \renewcommand\transparent[1]{}%
  }%
  \providecommand\rotatebox[2]{#2}%
  \newcommand*\fsize{\dimexpr\f@size pt\relax}%
  \newcommand*\lineheight[1]{\fontsize{\fsize}{#1\fsize}\selectfont}%
  \ifx\svgwidth\undefined%
    \setlength{\unitlength}{216.48304376bp}%
    \ifx\svgscale\undefined%
      \relax%
    \else%
      \setlength{\unitlength}{\unitlength * \real{\svgscale}}%
    \fi%
  \else%
    \setlength{\unitlength}{\svgwidth}%
  \fi%
  \global\let\svgwidth\undefined%
  \global\let\svgscale\undefined%
  \makeatother%
  \begin{picture}(1,0.81050501)%
    \lineheight{1}%
    \setlength\tabcolsep{0pt}%
    \put(0,0){\includegraphics[width=\unitlength,page=1]{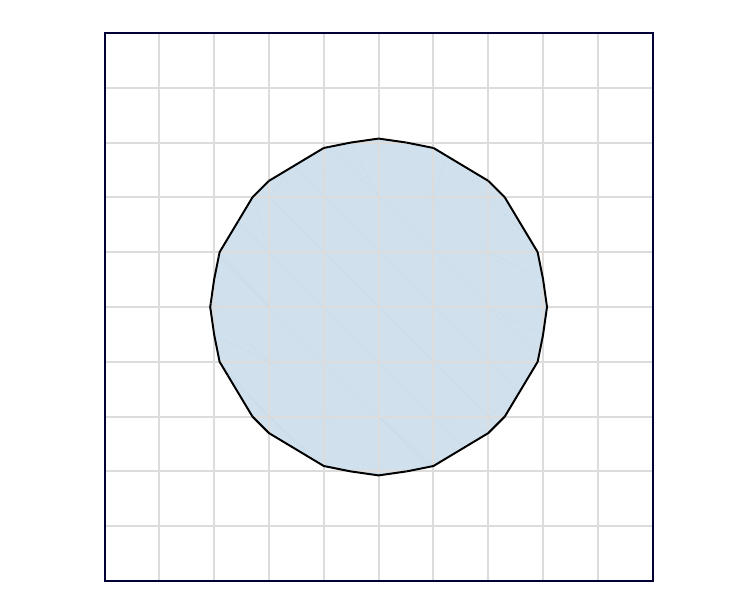}}%
    \put(0.49784789,0.42408172){\color[rgb]{0.39607843,0.06666667,0.12941176}\makebox(0,0)[t]{\lineheight{0}\smash{\begin{tabular}[t]{c}$\Omega_{\rm in}$ \end{tabular}}}}%
    \put(0.08732546,0.42015062){\color[rgb]{0.39607843,0.06666667,0.12941176}\makebox(0,0)[t]{\lineheight{0}\smash{\begin{tabular}[t]{c}$\partial \Omega$ \end{tabular}}}}%
    \put(0.49784789,0.37300237){\color[rgb]{0,0,0}\makebox(0,0)[t]{\lineheight{0}\smash{\begin{tabular}[t]{c}$ \mathcal{R}(\boldsymbol{u},\boldsymbol{v}) $ \end{tabular}}}}%
    \put(0.68824824,0.65165372){\color[rgb]{0,0,0}\makebox(0,0)[t]{\lineheight{0}\smash{\begin{tabular}[t]{c}$\mathcal{R}^{\mathrm{stb}}(\boldsymbol{u},\boldsymbol{v})$ \end{tabular}}}}%
    \put(0.68824824,0.70273306){\color[rgb]{0.39607843,0.06666667,0.12941176}\makebox(0,0)[t]{\lineheight{0}\smash{\begin{tabular}[t]{c}$\Omega_{\rm out}$ \end{tabular}}}}%
    \put(0.46031869,0.14311419){\color[rgb]{0.39607843,0.06666667,0.12941176}\makebox(0,0)[t]{\lineheight{0}\smash{\begin{tabular}[t]{c}$\Gamma$ \end{tabular}}}}%
  \end{picture}%
\endgroup%

%% file: figures/pertubation.pdf_tex
%% Creator: Inkscape 1.1.2 (0a00cf5339, 2022-02-04), www.inkscape.org
%% Accompanies image file 'pertubation.pdf' (pdf, eps, ps)
\begingroup%
  \makeatletter%
  \providecommand\color[2][]{%
    \errmessage{(Inkscape) Color is used for the text in Inkscape, but the package 'color.sty' is not loaded}%
    \renewcommand\color[2][]{}%
  }%
  \providecommand\transparent[1]{%
    \errmessage{(Inkscape) Transparency is used (non-zero) for the text in Inkscape, but the package 'transparent.sty' is not loaded}%
    \renewcommand\transparent[1]{}%
  }%
  \providecommand\rotatebox[2]{#2}%
  \newcommand*\fsize{\dimexpr\f@size pt\relax}%
  \newcommand*\lineheight[1]{\fontsize{\fsize}{#1\fsize}\selectfont}%
  \ifx\svgwidth\undefined%
    \setlength{\unitlength}{216.0683315bp}%
    \ifx\svgscale\undefined%
      \relax%
    \else%
      \setlength{\unitlength}{\unitlength * \real{\svgscale}}%
    \fi%
  \else%
    \setlength{\unitlength}{\svgwidth}%
  \fi%
  \global\let\svgwidth\undefined%
  \global\let\svgscale\undefined%
  \makeatother%
  \begin{picture}(1,0.80311247)%
    \lineheight{1}%
    \setlength\tabcolsep{0pt}%
    \put(0,0){\includegraphics[width=\unitlength,page=1]{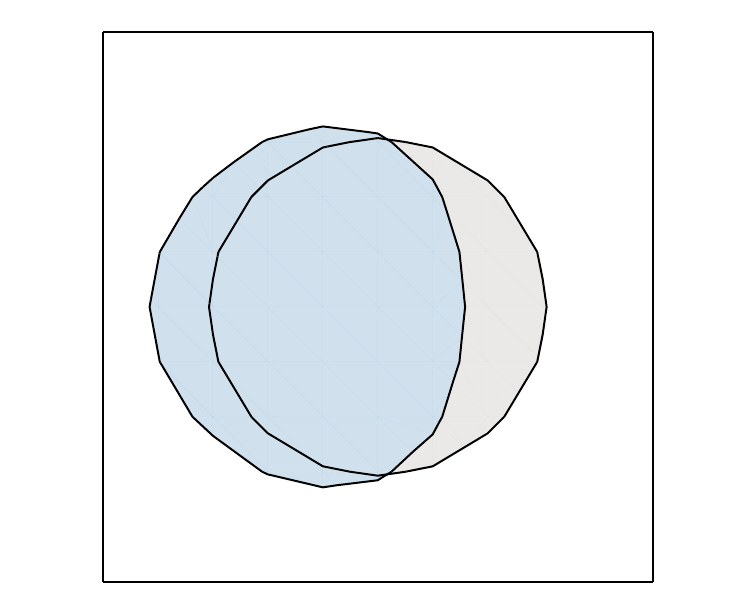}}%
    \put(0.42334554,0.38037393){\color[rgb]{0,0,0}\makebox(0,0)[t]{\lineheight{0}\smash{\begin{tabular}[t]{c}$\Omega_{\rm in}^\epsilon$ \end{tabular}}}}%
    \put(0.67462518,0.4137291){\color[rgb]{0,0,0}\makebox(0,0)[t]{\lineheight{0}\smash{\begin{tabular}[t]{c}$\epsilon$\end{tabular}}}}%
    \put(0,0){\includegraphics[width=\unitlength,page=2]{pertubation.pdf}}%
  \end{picture}%
\endgroup%

%% file: figures/GhostSkeleton.pdf_tex
%% Creator: Inkscape 1.1.2 (0a00cf5339, 2022-02-04), www.inkscape.org
%% Accompanies image file 'GhostSkeleton.pdf' (pdf, eps, ps)
\begingroup%
  \makeatletter%
  \providecommand\color[2][]{%
    \errmessage{(Inkscape) Color is used for the text in Inkscape, but the package 'color.sty' is not loaded}%
    \renewcommand\color[2][]{}%
  }%
  \providecommand\transparent[1]{%
    \errmessage{(Inkscape) Transparency is used (non-zero) for the text in Inkscape, but the package 'transparent.sty' is not loaded}%
    \renewcommand\transparent[1]{}%
  }%
  \providecommand\rotatebox[2]{#2}%
  \newcommand*\fsize{\dimexpr\f@size pt\relax}%
  \newcommand*\lineheight[1]{\fontsize{\fsize}{#1\fsize}\selectfont}%
  \ifx\svgwidth\undefined%
    \setlength{\unitlength}{216.56693346bp}%
    \ifx\svgscale\undefined%
      \relax%
    \else%
      \setlength{\unitlength}{\unitlength * \real{\svgscale}}%
    \fi%
  \else%
    \setlength{\unitlength}{\svgwidth}%
  \fi%
  \global\let\svgwidth\undefined%
  \global\let\svgscale\undefined%
  \makeatother%
  \begin{picture}(1,0.4662958)%
    \lineheight{1}%
    \setlength\tabcolsep{0pt}%
    \put(0,0){\includegraphics[width=\unitlength,page=1]{GhostSkeleton.pdf}}%
    \put(0.1435913,0.1158899){\color[rgb]{0,0,0}\makebox(0,0)[t]{\lineheight{1.25}\smash{\begin{tabular}[t]{c}\makebox[0pt]{$\Omega_{\rm in}$}\end{tabular}}}}%
    \put(0.35679224,0.32879503){\color[rgb]{0,0,0}\makebox(0,0)[lt]{\lineheight{1.25}\smash{\begin{tabular}[t]{l}\makebox[0pt]{$\Omega_{\rm out}$}\end{tabular}}}}%
    \put(0.11733643,0.27755193){\color[rgb]{0,0,0}\makebox(0,0)[lt]{\lineheight{1.25}\smash{\begin{tabular}[t]{l}$\Gamma$\end{tabular}}}}%
    \put(0.03858622,0.03672642){\color[rgb]{0,0,0}\makebox(0,0)[lt]{\lineheight{1.25}\smash{\begin{tabular}[t]{l}$\Gamma_G$\end{tabular}}}}%
    \put(0.92177538,0.39974169){\color[rgb]{0,0,0}\makebox(0,0)[lt]{\lineheight{1.25}\smash{\begin{tabular}[t]{l}$\Gamma_G^\epsilon$\end{tabular}}}}%
    \put(0.66901898,0.11510461){\color[rgb]{0,0,0}\makebox(0,0)[lt]{\lineheight{1.25}\smash{\begin{tabular}[t]{l}\makebox[0pt]{$\Omega_{\rm in}^\epsilon$}\end{tabular}}}}%
    \put(0.88024305,0.32868365){\color[rgb]{0,0,0}\makebox(0,0)[lt]{\lineheight{1.25}\smash{\begin{tabular}[t]{l}\makebox[0pt]{$\Omega_{\rm out}^\epsilon$}\end{tabular}}}}%
    \put(0.67003446,0.29259492){\color[rgb]{0,0,0}\makebox(0,0)[lt]{\lineheight{1.25}\smash{\begin{tabular}[t]{l}$\Gamma^\epsilon$\end{tabular}}}}%
    \put(0,0){\includegraphics[width=\unitlength,page=2]{GhostSkeleton.pdf}}%
  \end{picture}%
\endgroup%

%% file: derivative.tex
\section{Gradient Implementation}\label{backwards-pass-implementation}

To the best of our knowledge, there is no existing implementation of an unfitted \ac{ls} \ac{to} method that accepts arbitrary residuals defining the PDE and computes the entire gradient $\frac{dJ}{d\mathbf{p}}$ by automatic differentiation. Making use of a backward pass, we do this efficiently by defining differentiation rules for each of the steps in the method. 

\subsection{Integral Differentiation Operator}
The backward pass is mainly composed of gradients of integrals with respect to the  \ac{DOF} of \ac{fe} functions. To make the derivative computation efficient, we exploit the fact that the \ac{DOF} only have an effect on surrounding cells and utilize the optimizations exploiting sparsity in the \ac{fe} library Gridap \cite{Verdugo2021}. Integrals in the domain can be divided into cell-wise components:
\begin{equation}
\mathscr{I}(\boldsymbol{u},\boldsymbol{v},\phi) = \sum_{K\in\mathcal{T}_h} \mathscr{I}^K(\mathbf{u}^K,\mathbf{v}^K,\mathbf{\phi}^K), 
\end{equation}
where $\mathbf{u}^K \in \mathbb{R}^{\Sigma_u},\mathbf{v}^K \in \mathbb{R}^{\Sigma_v}$ and $\phi^K\in\mathbb{R}^{\Sigma_\phi}$ are the \ac{DOF} parameterizing the restrictions of $u,v$ and $\phi$ to the cell $K$ and $\Sigma u$, $\Sigma v$ and $\Sigma \phi$ are the number of \ac{DOF} in $K$ for the respective functions. 
Gradients can then be computed at roughly the cost of an integral evaluation for each cell $K$:
\begin{equation}
		\frac{\partial \mathscr{I} }{ \partial \phi }^K = \nabla^F_{\phi} \mathscr{I}^K ( \mathbf{u}^K,\mathbf{v}^K, \phi^K ),
\end{equation}
where the operator $\nabla^F_{\phi}$ represents taking the gradient with respect to $\phi^K$ using a vectorized forward propagation of dual numbers \cite{ForwardDiff}. 
To make taking derivatives in this way possible for the \ac{ls}, we implement the integrals so that each $\phi^K$ is accepted as the argument to compute the contribution $\mathscr{I}^K$:
\begin{equation}
		\mathscr{I}^K: \phi^K \in \mathbb{R}^{\Sigma_\phi} \mapsto \mathscr{I}_K(\mathbf{u}^K,\mathbf{v}^K,\phi^K) \in \mathbb{R}.
\end{equation}
where
\begin{equation}
		\mathscr{I}^K(\mathbf{u}^K,\mathbf{v}^K,\phi^K)=\int_{K(\phi^K)} \mathcal{I}(\mathbf{u}^K,\mathbf{v}^K)dK
\end{equation}

A key point is that the integral function subroutines, including all the unfitted \ac{fe} tools, are implemented in such a way as to allow the propagation of dual numbers through the code. We also make use of a reverse mode operator $\nabla^R$ for the backwards propagation of derivatives used where appropriate, e.g., for the \ac{nn}.

\subsection{Backwards Pass Routine}
We now present the backward pass in detail. 
To compute the sensitivity of the objective with respect to the parameters, we start with the seed $\frac{dJ}{dJ}=1$ and propagate derivatives in reverse mode using the chain rule:
\begin{equation}
		\frac{dJ}{d\mathbf{p}} = 	
		\frac{dJ}{dJ} \left(
		\frac{\partial{J}}{\partial{\phi }} +
		\frac{\partial{J}}{\partial{\boldsymbol{u}}} 
\frac{d\boldsymbol{u}}{d\phi } \right)
		\frac{d\phi }{d\mathbf{\varphi}} 
		\frac{d\mathbf{\varphi}}{d{\mathbf{p}}} 
\end{equation}
where an adjoint method on the problem residual $\mathscr{R}$ is used to differentiate through the PDE:
\begin{equation}
		\frac{\partial{J}}{\partial{\boldsymbol{u}}} \frac{d\boldsymbol{u}}{d\phi }  =  {-\lambda^T\frac{d\mathscr{R}}{d\phi }} \ (\text{ here we solved  } \frac{d\mathscr{R}}{d\boldsymbol{u}}^{T} \lambda =  \frac{dJ}{d\boldsymbol{u}}^{T}). 
\end{equation}
We then use the chain rule to differentiate through the \ac{ls} function processing steps:
\begin{equation}
		\frac{d\phi }{d\mathbf{\varphi}} = 
		\frac{d\phi }{d{\phi_{s(3)}}}  
		\frac{d{\phi_{s(3)}}}{d{\phi_{f(2)}}}  
		\frac{d{\phi_{f(2)}}}{d\mathbf{\varphi} }.
\end{equation}
The volume constraint here involved a root-finding method. To differentiate through this step, we utilize the implicit function theorem:
\begin{equation}
		\frac{d\phi }{d{\phi_{s(3)}}} =  \frac{\partial \phi } { \partial {\phi_{s(3)}}}  - \frac{\partial \phi  }{ \partial b} \frac{\partial \mathscr{V}}{\partial b}^{-1} \frac{\partial\mathscr{V }}{\partial {\phi_{s(3)}}},
\end{equation}
and to differentiate through the signed distance map, we use the adjoint method once again for the residual $\mathscr{R}_s$ equal to the integral in (\ref{eq:Rs}). 
Finally, we use standard backpropagation to compute the derivative with respect to the parameters of the \ac{nn}. The steps of the backward pass are presented explicitly in Algorithm \ref{al:bp}.

\begin{algorithm}
		\caption{Backwards Pass}\label{alg:cap}
		\begin{algorithmic}
		\State Initialize $\frac{dJ}{dJ} \gets 1$ 
		\State Extract $\phi^K \in \mathbb{R}^{\Sigma \phi}$, $\mathbf{u}^K \in \mathbb{R}^{\Sigma u}$ from $\phi$,$\boldsymbol{u}$ $\forall{K} \in \mathcal{T}_h$.
		\For{ $ K \in \mathcal{T}_h$ }
		\State $\frac{\partial J }{ \partial \boldsymbol{u} }^K \gets \nabla^F_{u} J ( \mathbf{u}^K, \phi^K )$ 
				\State $\frac{\partial J }{ \partial \phi }^K \gets \nabla^F_{\phi} J ( \mathbf{u}^K, \phi^K )$
		\EndFor
		\State Assemble the gradients $ \frac{\partial J }{ \partial \boldsymbol{u} } \in \mathbb{R}^{N_u}$ and $ \frac{\partial J }{ \partial \phi } \in \mathbb{R}^{N}$

		\State Assemble the sparse jacobian associated with the residual $ \frac{\partial \mathscr{R} }{ \partial \boldsymbol{u} } \in \mathbb{R}^{N_u,N_u}$
		\State Solve the adjoint equation $\frac{\partial \mathscr{R} }{ \partial \boldsymbol{u} } \lambda = \frac{\partial J }{ \partial \boldsymbol{u} }$ for $\lambda\in\mathbb{R}^{N_u}$
		\State Extract $\lambda^K \in \mathbb{R}^{\Sigma u}$  from $\lambda$ $\forall  K \in \mathcal{T}_h$ 
		\For{ $ K \in \mathcal{T}_h$ }	
		\State $\frac{\partial J }{ \partial \boldsymbol{u} }\frac{\partial \boldsymbol{u}}{\partial \phi}^K \gets \nabla^F_{u} \mathscr{R} ( \mathbf{u}^K, \lambda^K, \phi^K )$ 
		\EndFor
		\State Assemble the gradient $ \frac{\partial J }{ \partial \boldsymbol{u} }\frac{\partial \boldsymbol{u}}{\partial \phi} \in \mathbb{R}^N $ 
\State $\frac{dJ}{d\phi} \gets \frac{\partial J }{ \partial \phi } +  \frac{\partial J }{ \partial \boldsymbol{u} }\frac{\partial \boldsymbol{u}}{\partial \phi} $
		\State Compute the vector-jacobian-products: 
				\State $ \frac{\partial J }{ \partial b } \gets \frac{d J }{ d \phi } \nabla^R_b \phi  (\phi_{s(3)},b)  $ 
				\State $ \frac{\partial J }{ \partial \phi } \gets \frac{d J }{ d \phi } \nabla^R_{\phi} \phi  (\phi_{s(3)},b) $ 
		%\For{$ K \in \mathcal{T}_h$ }
		\State Compute the gradients:
				\State $\frac{\partial V }{ \partial \phi } \gets \nabla^R_{u} V ( \phi_{s(3)},b )$ 
				\State $\frac{\partial V }{ \partial b } \gets \nabla^F_{u} V ( \phi_{s(3)},b )$ 
		\State $\frac{dJ }{d{\phi_{s(3)}}} \gets \frac{\partial J } { \partial {\phi_{s(3)}}}  - \frac{\partial J  }{ \partial b} \frac{\partial \mathscr{V}}{\partial b}^{-1} \frac{\partial\mathscr{V }}{\partial {\phi_{s(3)}}}$
		
		\State Assemble the sparse jacobian associated with the residual $ \frac{\partial \mathscr{R}_s }{ \partial \phi_{s(3)} } \in \mathbb{R}^{N,N}$
		\State Solve the adjoint equation $\frac{\partial \mathscr{R}_s }{ \partial \phi_{s(3)} } \lambda_s = \frac{\partial J }{ \partial \phi_{s(3)} }$ for $\lambda_s\in\mathbb{R}^N$
		\For{ $ K \in \mathcal{T}_h$  }\\
				$\frac{d \mathscr{R}_s }{ d \phi_{f (2)} }^K \gets \nabla^F_{\phi_{f (2)} } \mathscr{R}_s ( \phi_{s(3)}^K, \lambda_s^K, \phi_{ f(2)}^K )$ 
		\EndFor
		\State Assemble the gradient $ \frac{d J }{ d \phi_{f(2)} }  \in \mathbb{R}^N$ 

		\State Compute the vector-jacobian-products: 
		\State $ \frac{d J }{d \varphi } \gets \frac{d J }{ d \phi_{f(2)} } \nabla^R_\varphi (\phi_{f(2)}(\varphi)) $ 
		
		\State $ \frac{d J }{d p } \gets \frac{d J }{ d \varphi } \nabla^R_p (N(\mathbf{p})) $ 
		\end{algorithmic}
		\label{al:bp}
\end{algorithm}

%% file: results.tex
\hypertarget{numerical-experiments}{%
\section{Numerical Experiments}\label{numerical-experiments}}

\hypertarget{benchmark-results}{%
\subsection{Benchmark Results}\label{benchmark-results}}

We first compare the optimized results obtained for benchmark problems against baseline methods using the model problems presented in Section \ref{simulation}.

The method presented in this work, the NN-LS method, is compared against its non-neural counterpart the Pixel \ac{ls} (Pixel-LS) method, that is, the same method without a neural prior where instead the nodal values of the level set function are taken as the optimization parameters. We also compare the method against the SIMP method of \ac{to} again using both a neural prior (NN-SIMP) and the standard approach (Pixel-SIMP). 
Following a standard SIMP implementation of the heat conduction problem, we use a conductivity based on the power law $k=\alpha_T + (1-\alpha_T)\rho^{\gamma}$ where $\rho$ is a design variable given by a \ac{fe} function constructed on the space $V_h^1$ where the \ac{DOF} values are the optimization parameters or output vector of the \ac{nn} in the Pixel-SIMP and NN-SIMP cases, respectively, and $\gamma$ is the penalization parameter, taken to be equal to $3$. 
Similarly, for the SIMP implementation of the structural problem, we use a Youngs Modulus $E=\alpha_d + (1- \alpha_T)\rho^{\gamma}$. 
To maintain a fair comparison between methods, we use the optimizers most widely regarded as suitable for the particular parameterization. Namely, we take the most commonly used MMA optimization strategy \cite{Svanberg1987} for the pixel parameterization and the ADAM strategy \cite{Kingma2014} for the \ac{nn}.

\subsubsection{Benchmark Problems}

\begin{figure}%
    \centering
	\begin{subfigure}[t]{0.49\textwidth}
		\centering\captionsetup{width=.9\linewidth}%
    \def\svgwidth{1\linewidth}
    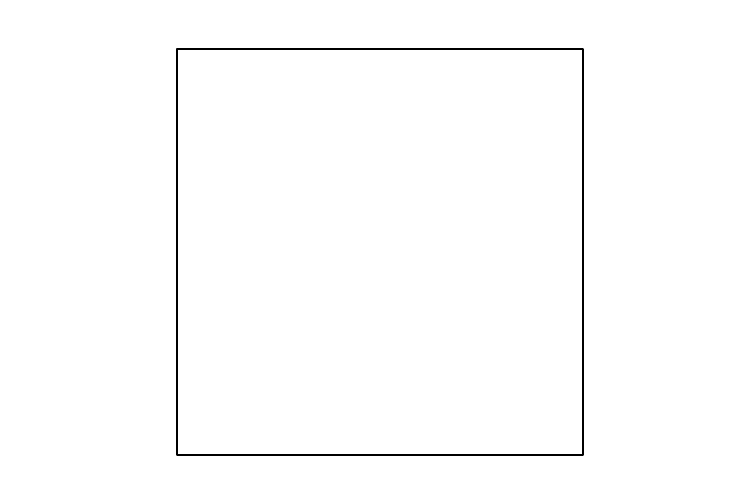

		\caption{Heat conduction problem setup. The top and left sides $\Gamma_D$ are given a Dirichlet condition and the bottom and left sides $\Gamma_N$ are given a Neumann condition.}
		\label{fig:heat-conduction-problem-setup}
	\end{subfigure}
	\begin{subfigure}[t]{0.49\textwidth}
		\centering\captionsetup{width=.9\linewidth}%
    \def\svgwidth{1\linewidth}
    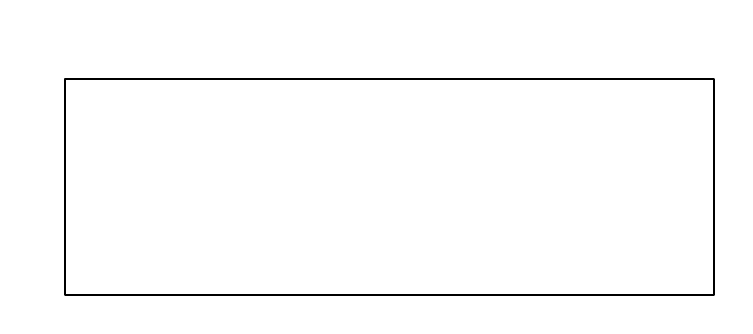

		\caption{The right half of the MBB problem exploiting symmetry. The roller supports provide vertical restraint on the left-hand side and horizontal restraint in the bottom right corner. A downward force $F$ is prescribed on the top left corner.}
		\label{fig:MBB-problem-setup}
	\end{subfigure}
    \caption{Benchmark problems}
    \label{fig:benchmark-problems}
\end{figure}

The first problem studied is the Poisson equation to model heat conduction with the setup in Figure \ref{fig:heat-conduction-problem-setup} selected from \cite{GersborgHansen2006}. For this problem, we set $k_0 =1 m^2s^{-1}$, $\alpha_T=0.01$, $f=0.01Ks^{-1}$, use homogenous Dirichlet and Neumann conditions and use a mesh of $ 95 \times 95 $ with a $0.4$ volume fraction. For the \ac{nn}, we set the number of convolutional layers to $5$, $N_{\boldsymbol{\Theta}} = 64$, $w = (12,12,24,46,96,96)$, $l = (12,12,24,46,96,96)$ and $c = (16 , 128, 64, 32, 16, 1 )$.  %\hypertarget{implementation-aspects}{%

The second problem studied is the typical MBB problem described in \cite{Sigmund2001} with the setup as in Figure \ref{fig:MBB-problem-setup}. For this problem, we set $\nu=0.3$, $E=1 Pa$, $\alpha_d = 0.001$ and $F=1 N$ and use a mesh of $ 287 \times 95 $ with a $0.4$ volume fraction.  We use the same network as in the Poisson problem but set $w = (36,36,64,128,256,256)$

\sbcom{
The hyperparameters of the network are mainly selected to match those presented in \cite{Hoyer2019} with minor adjustments for the specific problem. The size of the convolutional filter chosen is appropriate for this scale of problem since we have achieved the required level of expressivity, as can be seen in the subsequent results, through learning complex enough templates whilst retaining a reasonable number of parameters. 
}

\subsubsection{Optimized Structures}
To compare the SIMP and \ac{ls} results, the optimized densities \sbcom{ from the SIMP procedure are converted to an \ac{ls} by taking the 0.5 iso-surface of the density and recomputing the objective function value using the unfitted machinery which integrates exactly on the cut cells. To evaluate the effect of the \ac{nn} parameterization, we also include the results obtained through the more standard methods of parameterization by taking the coefficients of the \ac{fe} function defining the geometry directly as the design variables.
}

\begin{table*}%[h!]
	\begin{center}
	  %\caption{Method comparison}
	  \label{fig:comparison}
	  \begin{tabular}{c c c c c } %c c c c }
	    & NN-LS & Pixel-LS & NN-SIMP & Pixel-SIMP \\
		    Heat &
	    \raisebox{-.5\height}{\includegraphics[width=0.2\textwidth]{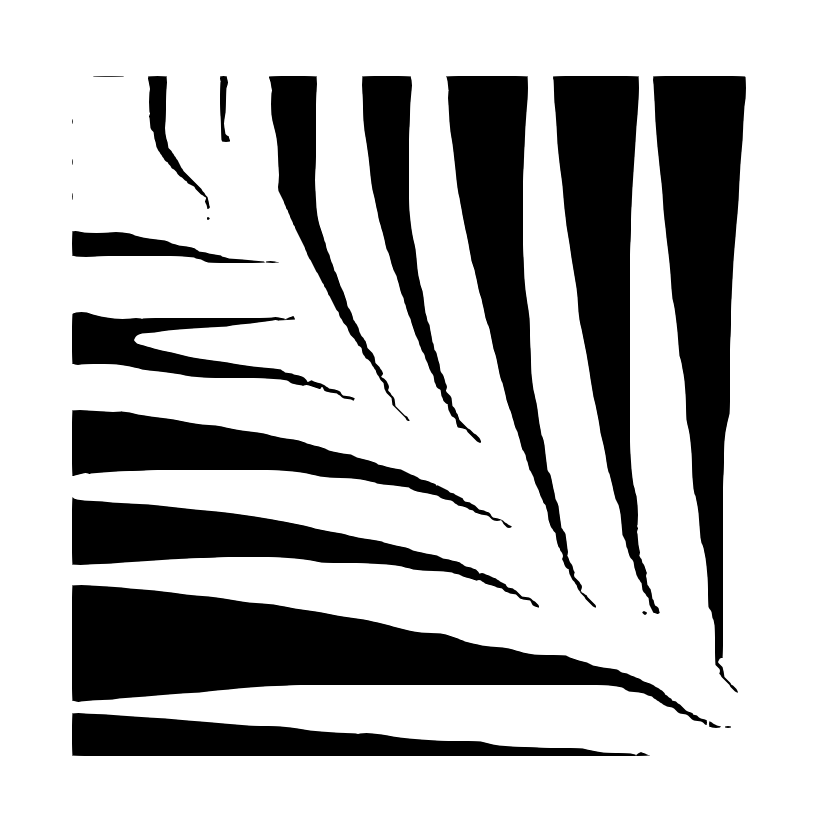}} & 
	    \raisebox{-.5\height}{\includegraphics[width=0.2\textwidth]{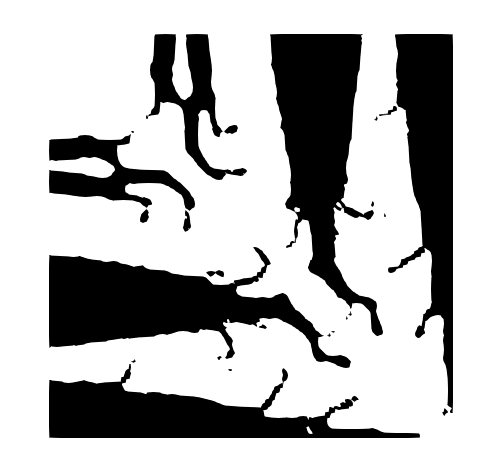}} &
	    \raisebox{-.5\height}{\includegraphics[width=0.2\textwidth]{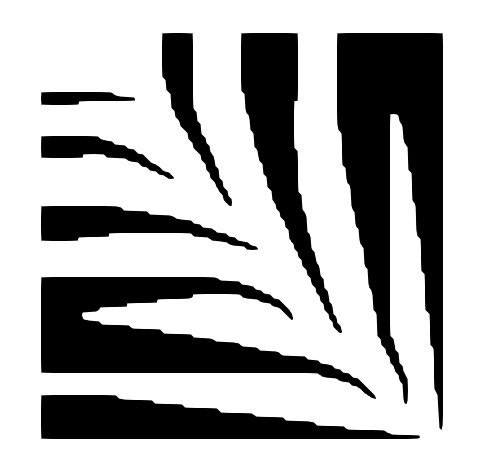}} &
	    \raisebox{-.5\height}{\includegraphics[width=0.2\textwidth]{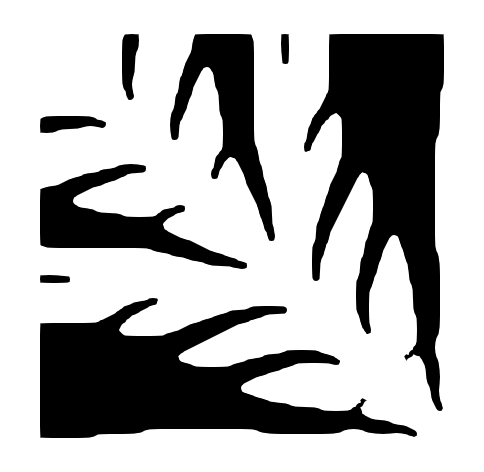}} \\ 
	    
	    &  0.0\% & 20.3\% & 5.8\% & 15.7\% \\
	    MBB  & 
	    \raisebox{-.5\height}{\includegraphics[width=0.2\textwidth]{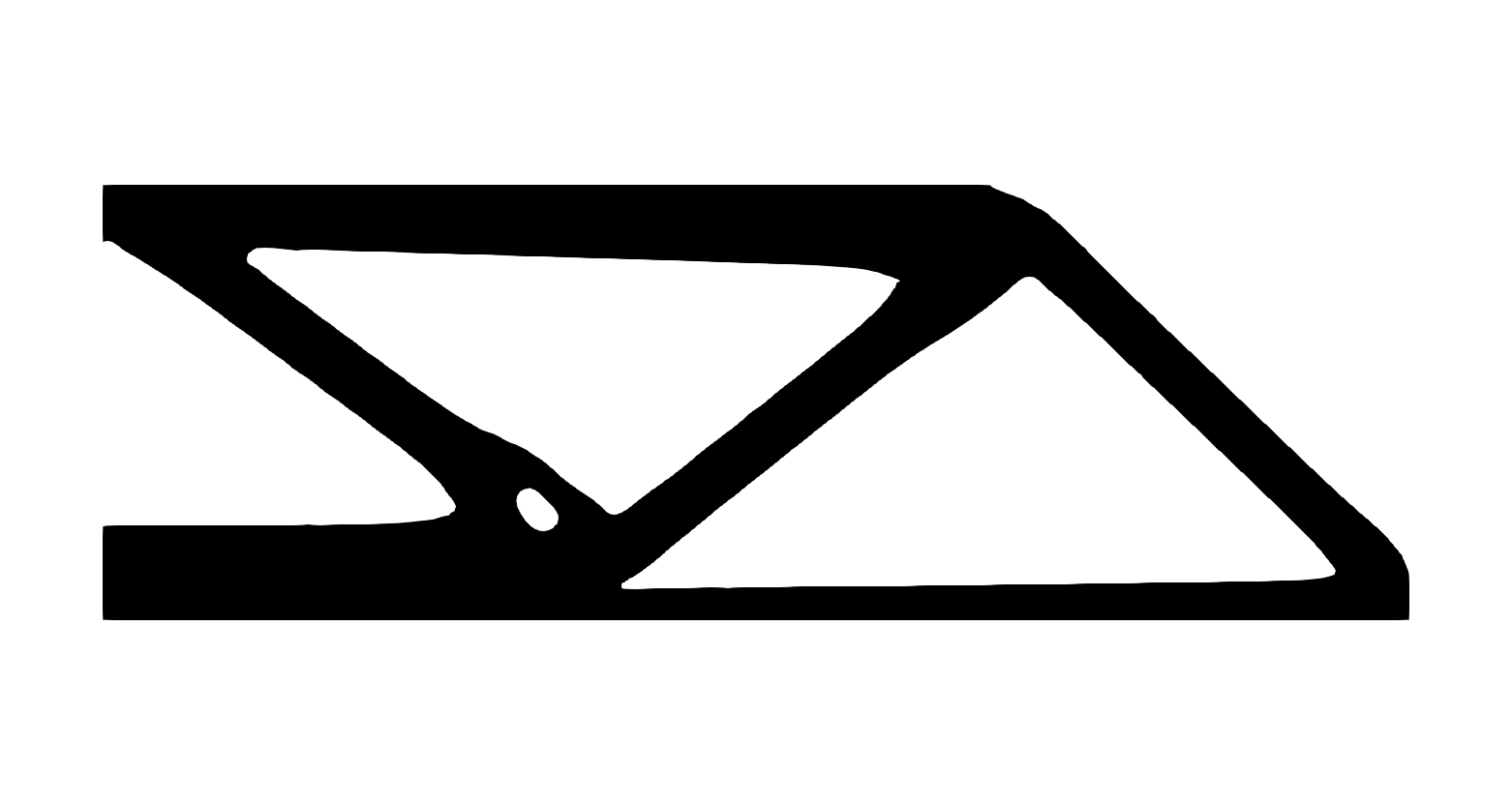}} & 
	    \raisebox{-.5\height}{\includegraphics[width=0.2\textwidth]{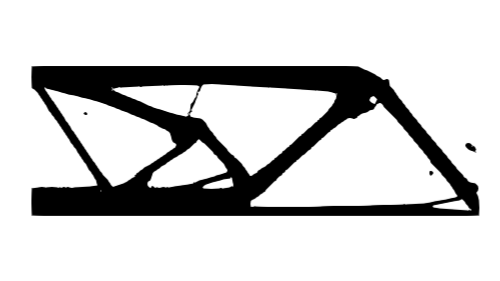}}&%P_LS_MBB.png}} &
	    \raisebox{-.5\height}{\includegraphics[width=0.2\textwidth]{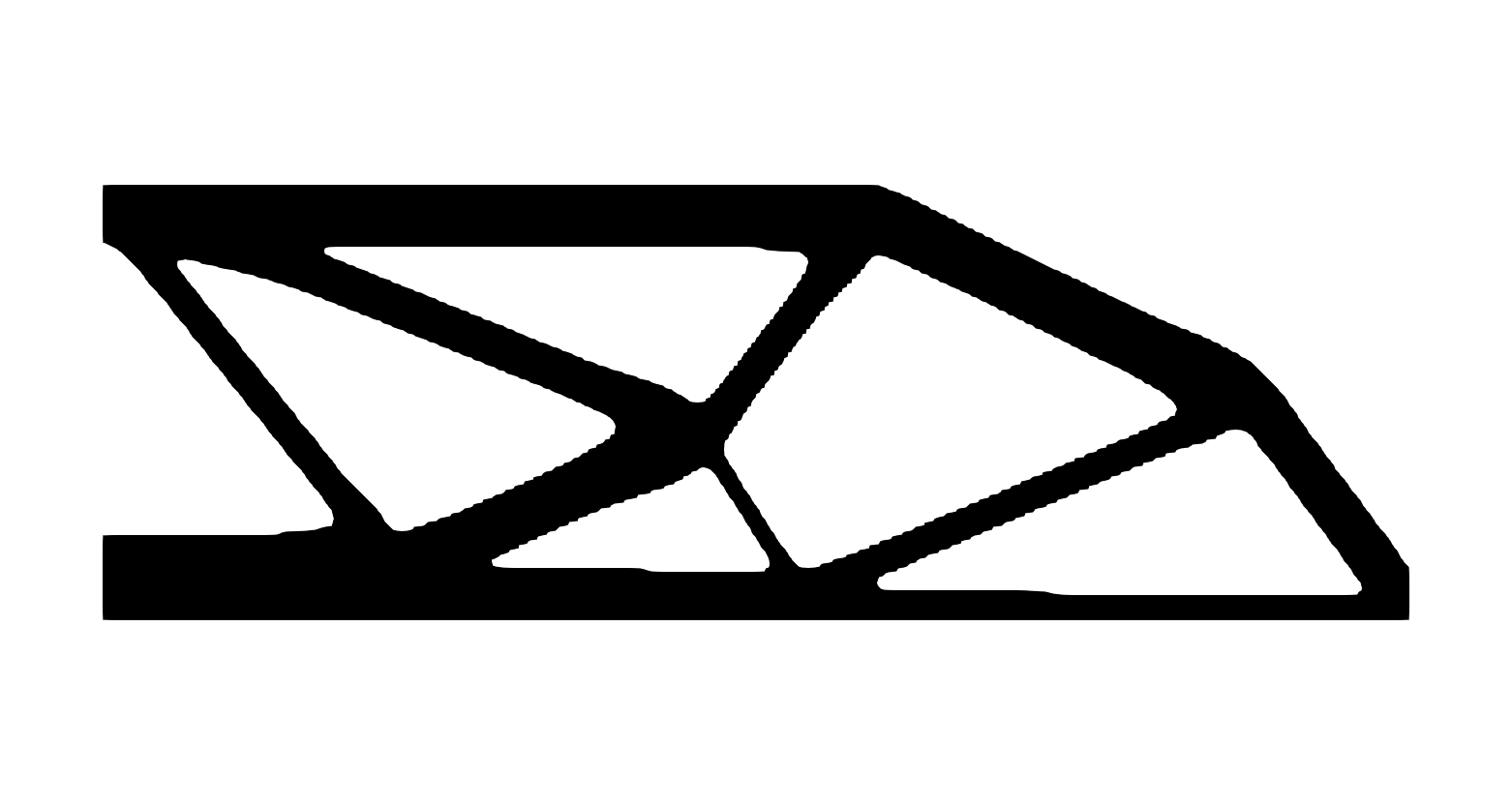}} &
	    \raisebox{-.5\height}{\includegraphics[width=0.2\textwidth]{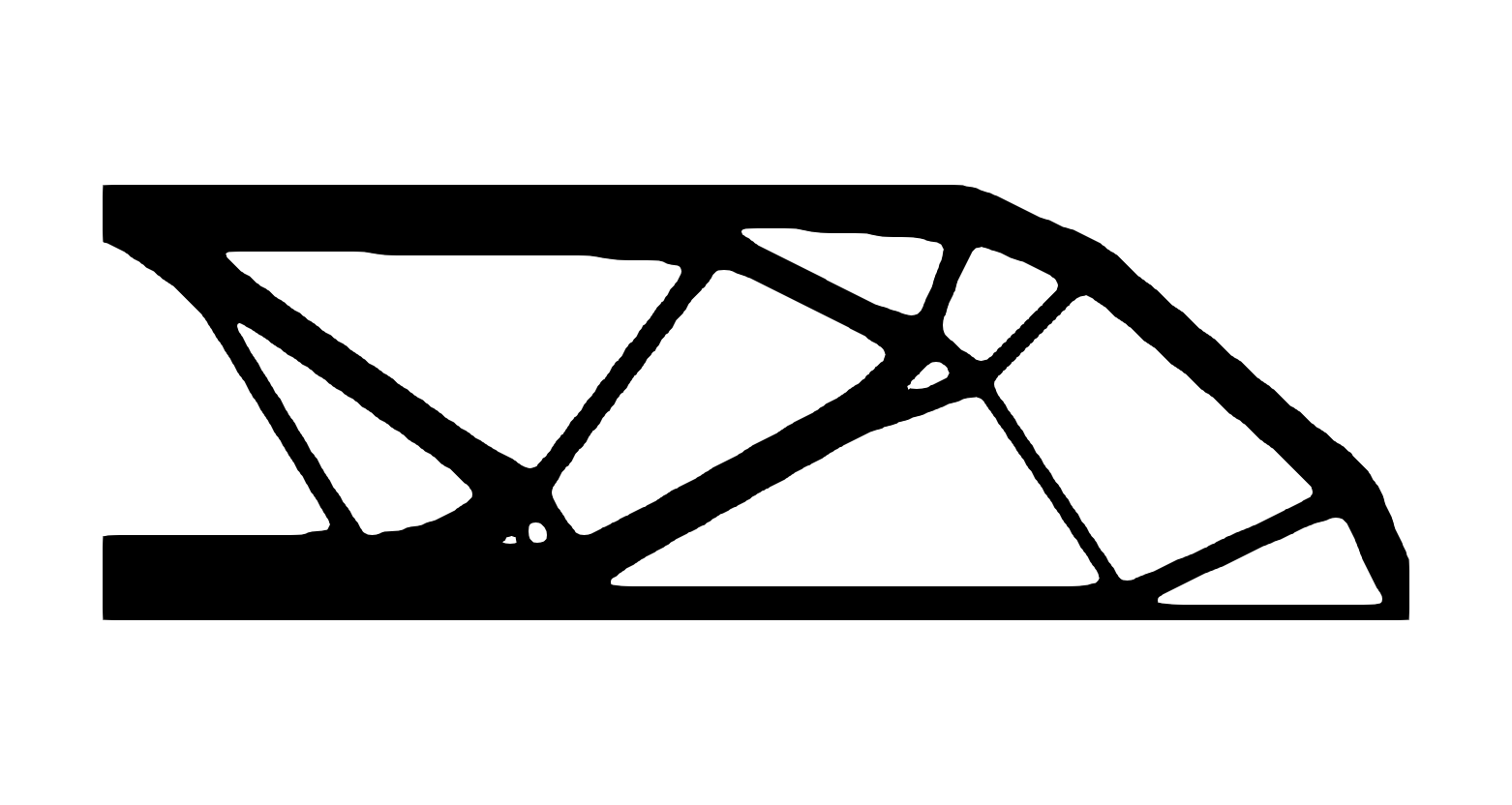}} \\ 
    
	    &  2.6\% & 1.7\% & 0.2\% & 0.0\% \\
    
	  \end{tabular}
	\end{center}
	\captionof{figure}{Optimized geometries for the various methods. The percentage under each geometry represents its performance relative to the best performing geometry in the row as measured by the objective function. \sbcom{The 0.0\% geometry is the best performing in the given row and the remainder are computed as the respective difference as compared to the best performing objective value.} } \label{table:mul-heat}
	\label{table:mul} 
      \end{table*}

The optimized geometries using the various methods are seen in Figure \ref{table:mul}. In all cases, the neural parameterization results in more regular geometries. The pixel-based methods could be regularized by augmenting the objective function with a penalization term although this would require manual tuning of a penalization parameter and may have an impact on the convergence. Simplistic structures could also be obtained for the pixel-based cases by controlling the filter radius and mesh resolution although this would prevent fine-scale structure. The NN-LS method instead allows for fine-scale features resolved in the final layer of the network but still produces performant regular geometries as the multi-scale influence of the parameters in the neural parameterization encourages globally performant structures to emerge. %occur rather than adding local features to a global structure fixed early in the optimization as seen most clearly in the comparison between the NN-SIMP and Pixel-SIMP results for the heat conduction problem in Figure \ref{fig:comparison} 
The use of the \ac{nn} is also seen to suppress numerical artifacts observed in the pixel-LS solutions. 
For the MBB problem, the compliance measurements for all methods fall within a few percent of each other. For this problem, the optimized solutions roughly share the location of the major members and are only slight variations from the optimal solution \cite{Rozvany1998}. 

For the heat conduction problem, however, the NN-LS method has the best performance by a fair amount followed by the NN-SIMP baseline. The regular geometries produced by the \ac{nn}s in these cases outperform their pixel counterparts by a significant amount in both the \ac{ls} and SIMP cases. The U-Net here seems able to find better minima because of its ability to focus on larger scale structures while the pixel-based parameterization makes improvements locally by adding finer scale branches to the geometry. Without handcrafting an initialisation close to the optimal solution, the addition of the neural network avoids sub-optimal branching structures and instead leads to the emergence of lamellar needles, the salient feature of optimal geometries for this problem \cite{Yan2018}.

\subsubsection{Convergence Plots}
\begin{figure}%
    \centering
    \begin{subfigure}{0.5\textwidth}
    \centering
		\includegraphics[width=0.99\linewidth]{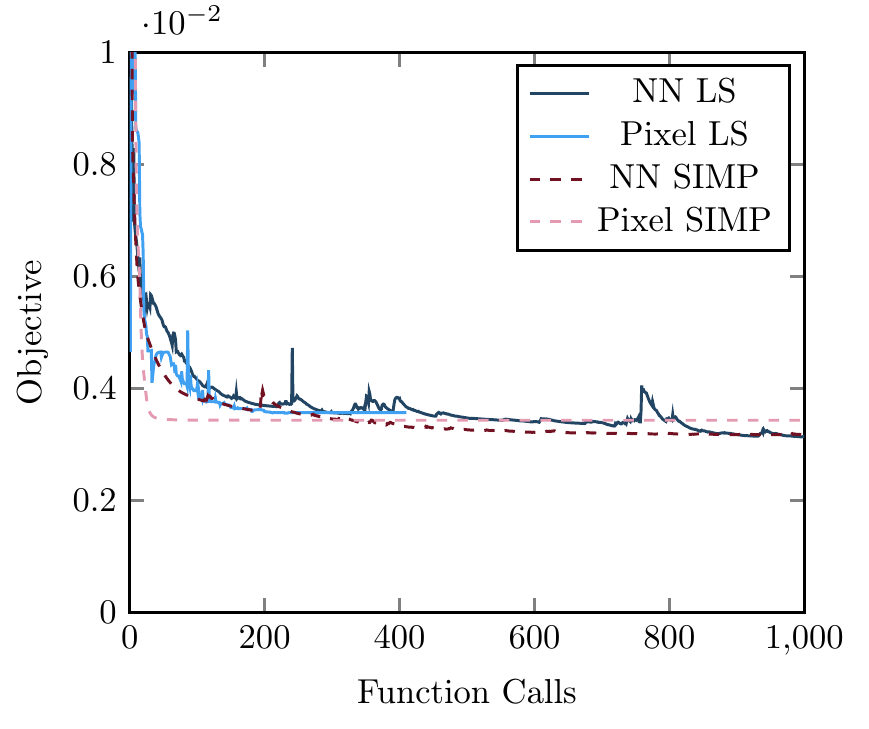} 
        \caption{Heat conduction}
        \label{fig:heat-convergnce}
    \end{subfigure}%
    \begin{subfigure}{0.5\textwidth}
    \centering
		\includegraphics[width=0.99\linewidth]{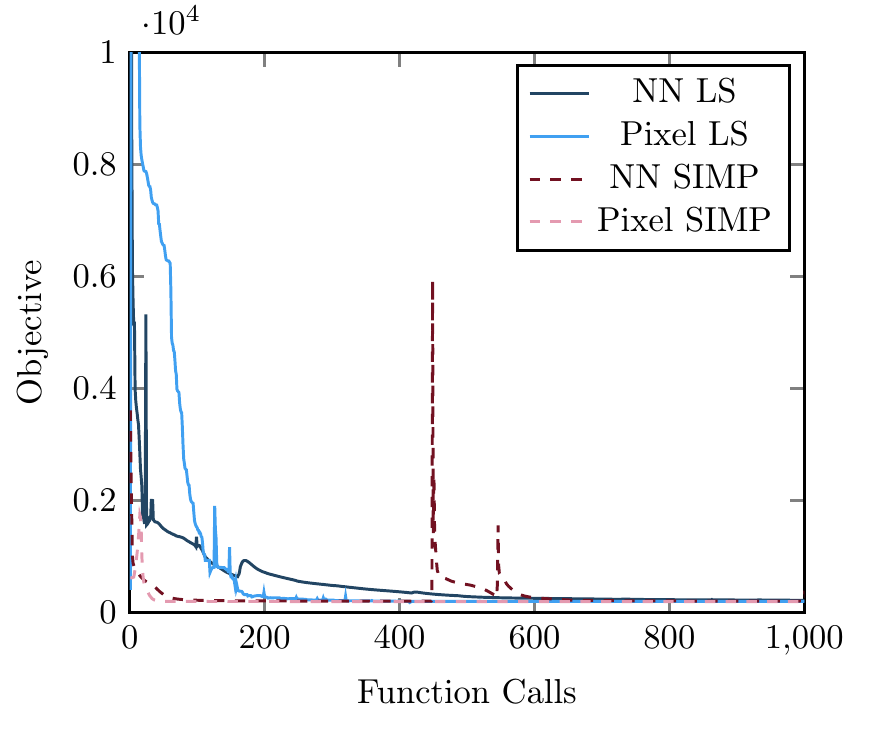} 
        \caption{MBB}
        \label{fig:MBB-convergence}
    \end{subfigure}
    \caption{Convergence plot comparison for the benchmark problems. The objective value at each objective function call is plotted. }
    \label{fig:convergence}
\end{figure}

The convergence of the methods is plotted for the heat conduction and MBB problems in Figure \ref{fig:convergence}. Since the use of the word iteration in the context of optimization is somewhat ambiguous, we plot the compliance against the number of objective function value calls for the process. For the ADAM and MMA optimizers, the ratio of function and gradient calls is 1:1.  
To compare the SIMP and \ac{ls} methods against each other, we bias the SIMP method using the optimized structures converted \ac{ls} once again. 
The bias is computed as the difference between the SIMP method's final objective value using the interpolated material and the SIMP method's final objective value using the converted \ac{ls}. This bias is applied to all of the series data in Figures \ref{fig:convergence} for the SIMP methods.   

For the heat conduction problem in Figure \ref{fig:heat-convergnce}, the pixel-SIMP method converges the fastest, albeit to an inferior solution. The rest of the methods converge at similar rates.  For the MBB problem in Figure \ref{fig:MBB-convergence}, the SIMP methods converge much faster than the \ac{ls} methods although the NN-SIMP method jumps out of the minimum at later iterations and stabilizes later on.

\hypertarget{FSI-problem}{%
\subsection{Interface Coupled Multiphysics Problem}\label{FSI-problem}}
\sbcom{
The design of a support in a fluid-structure problem is optimized in this section with the setup in \cite{Jenkins2016}. The structure is optimized under a forcing term from the fluid integrated on the evolving interface. The goal is to demonstrate the generality of the method which is shown here by its capacity to solve a multiphysics problem with interface coupling. Density methods do not extend naturally to handle such problems and are faced with difficulty in obtaining accurate coupling between the fluid and structure since the representation of the interface is spread across cells in the vicinity of the boundary. }

\begin{figure}
	\centering
    \def\svgwidth{0.55\linewidth}
    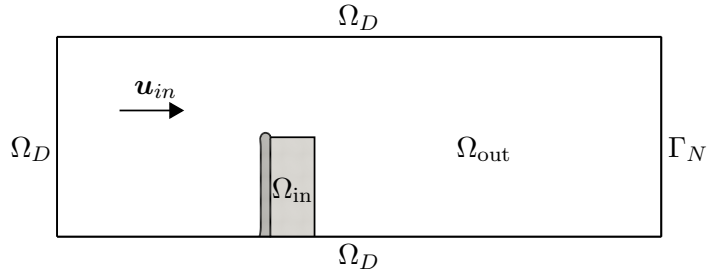

	\caption{The beam support problem setup. An inlet velocity is prescribed on the left side, a homogenous Dirichlet condition on the top and bottom walls and a homogenous Neumann condition on the right wall $\Omega_N$. 
	The design region supporting the beam in light grey can be either fluid or solid. 
	The solid domain $\Omega_{\mathrm{in}}$ is composed of the dark grey beam and the solid part of the design region.
	The fluid region $\Omega_{\mathrm{out}}$ is composed of the remainder of the channel, including the non-solid part of the design region.}
	\label{fig:FSISetup}
\end{figure}

\begin{figure}
		\centering
		\setlength{\abovecaptionskip}{-10pt plus 1pt minus 20pt }
    \def\svgwidth{1\linewidth}
    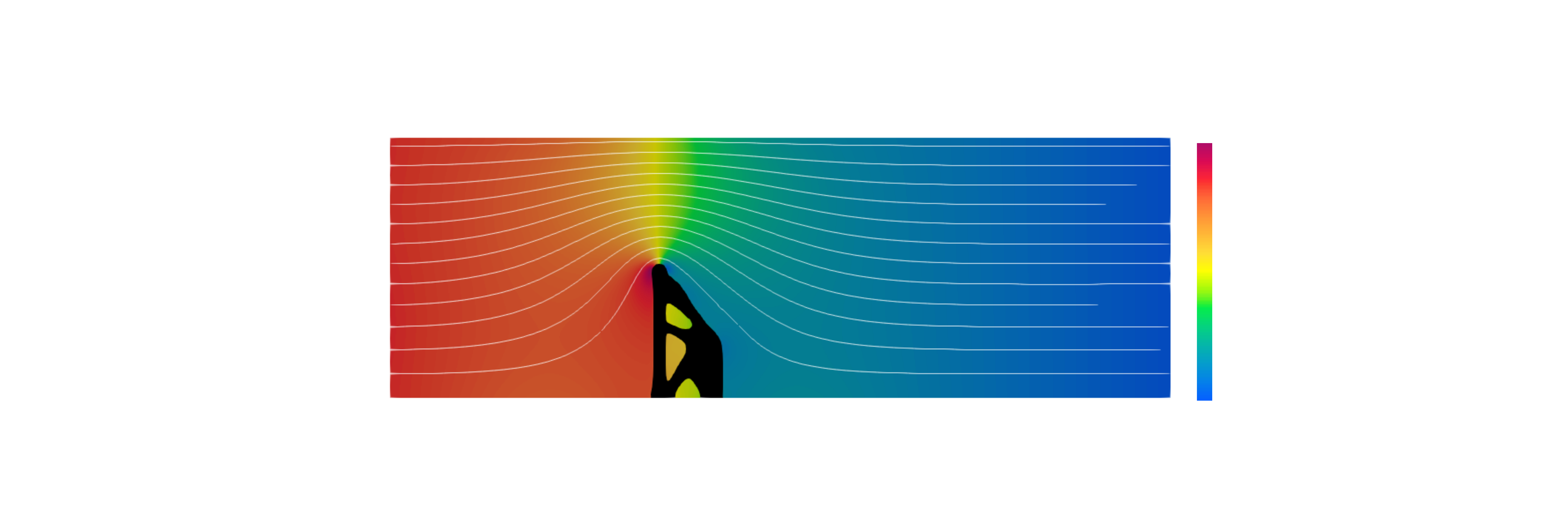
%}{fsi9del}
		\caption{Optimized geometry for the beam support problem. Streamlines are plotted to represent the velocity field and pressure is indicated in the legend. } 
		\label{fig:FSI-result}
		\end{figure}

The problem setup is seen in Figure \ref{fig:FSISetup}. 
For this problem, we set the fluid parameters as $\mu=1 m^2s^{-1}$ and $\alpha_u=2.5 \mu / 0.01^2 $ and use a parabolic velocity profile on the inlet with an average velocity of $0.01 ms^{-1}$.
For the structural parameters, we set $E=1 Pa$ and $\nu=0.3$ and use a $0.45$ volume fraction. 
We use the same network as in the poisson problem but set $w = (12,12,24,46,96,96)$ and $l = (24,24,48,96,192,192)$. The optimized geometry for problem is seen in Figure \ref{fig:FSI-result}.

%\begin{comment}
%\end{comment}

%% file: figures/heat_setup.pdf_tex
%% Creator: Inkscape 1.3-dev (74adff3496, 2022-05-26), www.inkscape.org
%% Accompanies image file 'heat_setup.pdf' (pdf, eps, ps)
\begingroup%
  \makeatletter%
  \providecommand\color[2][]{%
    \errmessage{(Inkscape) Color is used for the text in Inkscape, but the package 'color.sty' is not loaded}%
    \renewcommand\color[2][]{}%
  }%
  \providecommand\transparent[1]{%
    \errmessage{(Inkscape) Transparency is used (non-zero) for the text in Inkscape, but the package 'transparent.sty' is not loaded}%
    \renewcommand\transparent[1]{}%
  }%
  \providecommand\rotatebox[2]{#2}%
  \newcommand*\fsize{\dimexpr\f@size pt\relax}%
  \newcommand*\lineheight[1]{\fontsize{\fsize}{#1\fsize}\selectfont}%
  \ifx\svgwidth\undefined%
    \setlength{\unitlength}{216.24095526bp}%
    \ifx\svgscale\undefined%
      \relax%
    \else%
      \setlength{\unitlength}{\unitlength * \real{\svgscale}}%
    \fi%
  \else%
    \setlength{\unitlength}{\svgwidth}%
  \fi%
  \global\let\svgwidth\undefined%
  \global\let\svgscale\undefined%
  \makeatother%
  \begin{picture}(1,0.67045943)%
    \lineheight{1}%
    \setlength\tabcolsep{0pt}%
    \put(0,0){\includegraphics[width=\unitlength,page=1]{heat_setup.pdf}}%
    \put(0.51169611,0.32235784){\color[rgb]{0,0,0}\makebox(0,0)[t]{\lineheight{0}\smash{\begin{tabular}[t]{c}$\Omega$ \end{tabular}}}}%
    \put(0,0){\includegraphics[width=\unitlength,page=2]{heat_setup.pdf}}%
    \put(0.55240595,0.02125325){\color[rgb]{0,0,0}\makebox(0,0)[rt]{\lineheight{0}\smash{\begin{tabular}[t]{r}$\Gamma_{N}$ \end{tabular}}}}%
    \put(0.1673945,0.32297561){\color[rgb]{0,0,0}\makebox(0,0)[lt]{\lineheight{0}\smash{\begin{tabular}[t]{l}$\Gamma_{D}$ \end{tabular}}}}%
    \put(0.48374912,0.62100933){\color[rgb]{0,0,0}\makebox(0,0)[lt]{\lineheight{0}\smash{\begin{tabular}[t]{l}$\Gamma_{D}$ \end{tabular}}}}%
    \put(0.85207157,0.32297561){\color[rgb]{0,0,0}\makebox(0,0)[rt]{\lineheight{0}\smash{\begin{tabular}[t]{r}$\Gamma_{N}$ \end{tabular}}}}%
  \end{picture}%
\endgroup%

%% file: figures/MBB_setup.pdf_tex
%% Creator: Inkscape 1.1.2 (0a00cf5339, 2022-02-04), www.inkscape.org
%% Accompanies image file 'MBB_setup.pdf' (pdf, eps, ps)
\begingroup%
  \makeatletter%
  \providecommand\color[2][]{%
    \errmessage{(Inkscape) Color is used for the text in Inkscape, but the package 'color.sty' is not loaded}%
    \renewcommand\color[2][]{}%
  }%
  \providecommand\transparent[1]{%
    \errmessage{(Inkscape) Transparency is used (non-zero) for the text in Inkscape, but the package 'transparent.sty' is not loaded}%
    \renewcommand\transparent[1]{}%
  }%
  \providecommand\rotatebox[2]{#2}%
  \newcommand*\fsize{\dimexpr\f@size pt\relax}%
  \newcommand*\lineheight[1]{\fontsize{\fsize}{#1\fsize}\selectfont}%
  \ifx\svgwidth\undefined%
    \setlength{\unitlength}{216.61038136bp}%
    \ifx\svgscale\undefined%
      \relax%
    \else%
      \setlength{\unitlength}{\unitlength * \real{\svgscale}}%
    \fi%
  \else%
    \setlength{\unitlength}{\svgwidth}%
  \fi%
  \global\let\svgwidth\undefined%
  \global\let\svgscale\undefined%
  \makeatother%
  \begin{picture}(1,0.44403514)%
    \lineheight{1}%
    \setlength\tabcolsep{0pt}%
    \put(0,0){\includegraphics[width=\unitlength,page=1]{MBB_setup.pdf}}%
    \put(0.50540026,0.18517762){\color[rgb]{0,0,0}\makebox(0,0)[t]{\lineheight{0}\smash{\begin{tabular}[t]{c}$\Omega$ \end{tabular}}}}%
    \put(0,0){\includegraphics[width=\unitlength,page=2]{MBB_setup.pdf}}%
    \put(0.13777781,0.37835979){\color[rgb]{0,0,0}\makebox(0,0)[rt]{\lineheight{0}\smash{\begin{tabular}[t]{r}$F$ \end{tabular}}}}%
    \put(0,0){\includegraphics[width=\unitlength,page=3]{MBB_setup.pdf}}%
  \end{picture}%
\endgroup%

%% file: figures/FSI-setup.pdf_tex
%% Creator: Inkscape 1.3-dev (74adff3496, 2022-05-26), www.inkscape.org
%% Accompanies image file 'FSI-setup.pdf' (pdf, eps, ps)
\begingroup%
  \makeatletter%
  \providecommand\color[2][]{%
    \errmessage{(Inkscape) Color is used for the text in Inkscape, but the package 'color.sty' is not loaded}%
    \renewcommand\color[2][]{}%
  }%
  \providecommand\transparent[1]{%
    \errmessage{(Inkscape) Transparency is used (non-zero) for the text in Inkscape, but the package 'transparent.sty' is not loaded}%
    \renewcommand\transparent[1]{}%
  }%
  \providecommand\rotatebox[2]{#2}%
  \newcommand*\fsize{\dimexpr\f@size pt\relax}%
  \newcommand*\lineheight[1]{\fontsize{\fsize}{#1\fsize}\selectfont}%
  \ifx\svgwidth\undefined%
    \setlength{\unitlength}{216.01037213bp}%
    \ifx\svgscale\undefined%
      \relax%
    \else%
      \setlength{\unitlength}{\unitlength * \real{\svgscale}}%
    \fi%
  \else%
    \setlength{\unitlength}{\svgwidth}%
  \fi%
  \global\let\svgwidth\undefined%
  \global\let\svgscale\undefined%
  \makeatother%
  \begin{picture}(1,0.40687805)%
    \lineheight{1}%
    \setlength\tabcolsep{0pt}%
    \put(0,0){\includegraphics[width=\unitlength,page=1]{FSI-setup.pdf}}%
    \put(0.68841542,0.17588312){\color[rgb]{0,0,0}\makebox(0,0)[t]{\lineheight{0}\smash{\begin{tabular}[t]{c}$\Omega_{\mathrm{out}}$ \end{tabular}}}}%
    \put(0.50128452,0.3738776){\color[rgb]{0,0,0}\makebox(0,0)[t]{\lineheight{0}\smash{\begin{tabular}[t]{c}$\Omega_D$ \end{tabular}}}}%
    \put(0.01130698,0.17588312){\color[rgb]{0,0,0}\makebox(0,0)[t]{\lineheight{0}\smash{\begin{tabular}[t]{c}$\Omega_D$ \end{tabular}}}}%
    \put(0.50128452,0.01387477){\color[rgb]{0,0,0}\makebox(0,0)[t]{\lineheight{0}\smash{\begin{tabular}[t]{c}$\Omega_D$ \end{tabular}}}}%
    \put(0.4017045,0.11428698){\color[rgb]{0,0,0}\makebox(0,0)[t]{\lineheight{0}\smash{\begin{tabular}[t]{c}$\Omega_{\mathrm{in}}$ \end{tabular}}}}%
    \put(0.19703063,0.27096374){\color[rgb]{0,0,0}\makebox(0,0)[t]{\lineheight{0}\smash{\begin{tabular}[t]{c}$\boldsymbol{u}_{in}$ \end{tabular}}}}%
    \put(0.99330531,0.17588312){\color[rgb]{0,0,0}\makebox(0,0)[t]{\lineheight{0}\smash{\begin{tabular}[t]{c}$\Gamma_N$ \end{tabular}}}}%
    \put(0,0){\includegraphics[width=\unitlength,page=2]{FSI-setup.pdf}}%
  \end{picture}%
\endgroup%

%% file: figures/FSI_updated.pdf_tex
%% Creator: Inkscape inkscape 0.92.5, www.inkscape.org
%% Accompanies image file 'FSI_updated.pdf' (pdf, eps, ps)
\begingroup%
  \makeatletter%
  \providecommand\color[2][]{%
    \errmessage{(Inkscape) Color is used for the text in Inkscape, but the package 'color.sty' is not loaded}%
    \renewcommand\color[2][]{}%
  }%
  \providecommand\transparent[1]{%
    \errmessage{(Inkscape) Transparency is used (non-zero) for the text in Inkscape, but the package 'transparent.sty' is not loaded}%
    \renewcommand\transparent[1]{}%
  }%
  \providecommand\rotatebox[2]{#2}%
  \newcommand*\fsize{\dimexpr\f@size pt\relax}%
  \newcommand*\lineheight[1]{\fontsize{\fsize}{#1\fsize}\selectfont}%
  \ifx\svgwidth\undefined%
    \setlength{\unitlength}{1184bp}%
    \ifx\svgscale\undefined%
      \relax%
    \else%
      \setlength{\unitlength}{\unitlength * \real{\svgscale}}%
    \fi%
  \else%
    \setlength{\unitlength}{\svgwidth}%
  \fi%
  \global\let\svgwidth\undefined%
  \global\let\svgscale\undefined%
  \makeatother%
  \begin{picture}(1,0.33783784)%
    \lineheight{1}%
    \setlength\tabcolsep{0pt}%
    \put(0,0){\includegraphics[width=\unitlength,page=1]{FSI_updated.pdf}}%
    \put(0.78180223,0.08428389){\color[rgb]{0,0,0}\makebox(0,0)[lt]{\lineheight{1.25}\smash{\begin{tabular}[t]{l}0.0\end{tabular}}}}%
    \put(0.78180223,0.24838561){\color[rgb]{0,0,0}\makebox(0,0)[lt]{\lineheight{1.25}\smash{\begin{tabular}[t]{l}0.5\end{tabular}}}}%
    \put(0.78180223,0.11722627){\color[rgb]{0,0,0}\makebox(0,0)[lt]{\lineheight{1.25}\smash{\begin{tabular}[t]{l}0.1\end{tabular}}}}%
    \put(0.78180223,0.15016861){\color[rgb]{0,0,0}\makebox(0,0)[lt]{\lineheight{1.25}\smash{\begin{tabular}[t]{l}0.2\end{tabular}}}}%
    \put(0.78180223,0.18311094){\color[rgb]{0,0,0}\makebox(0,0)[lt]{\lineheight{1.25}\smash{\begin{tabular}[t]{l}0.3\end{tabular}}}}%
    \put(0.78180223,0.21605327){\color[rgb]{0,0,0}\makebox(0,0)[lt]{\lineheight{1.25}\smash{\begin{tabular}[t]{l}0.4\end{tabular}}}}%
    \put(0.86561221,0.16431526){\color[rgb]{0,0,0}\makebox(0,0)[t]{\lineheight{1.25}\smash{\begin{tabular}[t]{c}$p_h $[Pa]\end{tabular}}}}%
    \put(0,0){\includegraphics[width=\unitlength,page=2]{FSI_updated.pdf}}%
  \end{picture}%
\endgroup%

%% file: conclusion.tex
\hypertarget{conclusion}{%
\section{Conclusion}\label{conclusion}}
Here we propose a neural \ac{ls} method as a means of coupling convolutional \ac{nn}s and the unfitted \ac{ls} method. The neural parameterization provides more regular geometries and similar, in some cases better, performance compared to the well-known SIMP method of \ac{to} and the pixel-based counterpart \ac{ls} method.
The neural parameterization learns features at multiple scales during the optimization allowing for the emergence of regular structures whilst maintaining performance. 
The method takes longer to converge than the SIMP method and may not be suitable for simple problems such as those in linear elasticity. In this case, the network can be used in conjunction with the SIMP method. In contrast to SIMP, however, the method extends naturally to interface multiphysics problems and is more suitable in the general case. 
It should be noted that in our approach, the expressivity of geometries for the neural parameterization is the same as pixel counterpart methods since we use the same space $V_h^1$ in all cases to describe the geometry. A direction for development in the future would be to relax this constraint and also allow for the \ac{nn} to control the space in which it expresses itself, potentially allowing for more efficient expressions of better topologies. Further investigation into architecture design and optimization strategy may also lead to a greater and more general improvement in performance. 

\hypertarget{replication-of-results}{%
\section{Replication of Results}\label{replication-of-results}}

All the codes being used in this paper are distributed as open-source software. For reproducibility purposes, the implementation of the proposed \ac{to} methodology as well as the drivers being used to compare the different numerical methods are publicly available in the following repository: \url{https://github.com/ConnorMallon/NLSTO}

\section*{Acknowledgements}
\label{sect:acknowledgements}
This research was partially funded by the Australian Government through the Australian Research Council (project number DP220103160).
\section*{Conflict of Interest}
\label{sect:conflict-of-interest}
On behalf of all authors, the corresponding author states that there is no conflict of interest.